\documentclass[letterpaper,showpacs,showkeys,prc,twocolumn,nofootinbib]{revtex4}

\usepackage{graphicx}
\usepackage{amsmath}
\usepackage{amssymb}
\usepackage[sort&compress]{natbib}
\usepackage{ifthen}
\usepackage{hyperref} 

\newcommand{\infinity}{\infty}
\newcommand{\wv}[1]{\mathbf{#1}}
\newcommand{\lessim}{\lesssim}
\newcommand{\eq}[1]{Eq.~(\ref{#1})}

\newcommand{\fig}[1]{Fig.~\ref{#1}}

\newcommand{\tab}[1]{Table \ref{#1}}
\newcommand{\raa}{$R_{AA}$ }
\newcommand{\raacomma}{$R_{AA}$}

\newcommand{\raapt}{$R_{AA}(\eqnpt)$ }

\newcommand{\be}{\begin{equation}}
\newcommand{\ee}{\end{equation}}
\newcommand{\as}{\alpha_s}
\newcommand{\alphas}{$\as$ }

\newcommand{\pt}{$p_T$ }

\newcommand{\ptcomma}{$p_T$}

\newcommand{\eqnpt}{p_T}

\newcommand{\highpt}{high-\pt}

\newcommand{\wk}{\mathbf{k}}
\newcommand{\wq}{\mathbf{q}}

\newcommand{\kmax}{\mbox{$k_\mathrm{max}$}}
\newcommand{\thetamax}{\mbox{$\theta_\mathrm{max}$}}

\begin{document}

\title{Systematic Uncertainties in Theoretical Predictions of Jet Quenching}

\date{\today}

\author{W.\ A.\ Horowitz}
\email{horowitz@mps.ohio-state.edu}
\affiliation{Physics Department, The Ohio State University,\\191 West Woodruff Avenue, Columbus, OH 43210, USA}
\author{B.\ A.\ Cole}
\email{cole@nevis.columbia.edu}
\affiliation{Physics Department, Columbia University,\\538 West 120$^{th}$ Street, New York, NY 10027, USA}

\begin{abstract}
We find that the current radiative energy loss
kernels obtained from the opacity expansion dramatically violate the collinear approximation used in their 
derivation.
By keeping only the lowest order in collinearity terms, models based on the opacity expansion have $\sim50\%$ systematic uncertainty in the
calculation of $\pi^0$ $R_{AA}$ in \mbox{0-5\%} most central RHIC
collisions resulting in a systematic uncertainty of $\sim200\%$ in the
extracted medium density.  Surprisingly, the inclusion of a thermal
gluon mass on the order of the Debye screening scale affects $R_{AA}$
at only about the 5\% level due to non-intuitive coherence effects.
For some observables such as $R_{AA}$, the effect of these
uncertainties decreases with increasing jet energy; for others, such as
the average number of radiated gluons, the effect is energy
independent.  We note that it is likely that the differences reported
in the extracted values of medium parameters such as $\hat{q}$ by
various jet energy loss models will fall within this collinear
approximation systematic uncertainty; it is imperative for the
quantitative extraction of medium parameters or the possible
falsification of the hypothesis of weak coupling between the hard probes and soft modes of the quark gluon plasma medium that future radiative
energy loss research push beyond the lowest order collinear
approximation. 
\end{abstract}


\maketitle

\section{Introduction}
Jet quenching is a unique observable in ultra-relativistic heavy ion
collisions (URHIC) as high transverse momentum (high-\ptcomma)
particles are the most controlled, calibrated, and direct probe of the
fundamental soft degrees of freedom of the quark-gluon plasma (QGP) \cite{Gyulassy:2004zy}. Rigorous falsification or confirmation of fundamentally different
qualitative pictures of the basic physics of the QGP (e.g., whether it
is strongly or weakly coupled \cite{Horowitz:2007su}, its relevant degrees of freedom \cite{Gyulassy:1993hr,Herzog:2006gh,Gubser:2006bz}, etc.)\ from comparing theoretical predictions to \highpt data crucially
requires a detailed understanding of both experimental \emph{and
  theoretical} uncertainties. 

\begin{table}[!htbp]
\begin{tabular}{|c||c|c|c|c|c|}
 \hline
$x$: & \multicolumn{2}{|c|}{$x_+$} & $x_+(x_E)$ & \multicolumn{2}{|c|}{$x_E$} \\
 \hline
$\theta_\mathrm{max}:$ & \multicolumn{4}{|c|}{$\pi/2$} & $\pi/4$ \\
 \hline
$m_g$: &  $\mu/\surd2$ & \multicolumn{4}{|c|}{$0$} \\
 \hline\hline
\begin{tabular}{c} Rad Only \\ $dN_g/dy$ ($\times$ 1000) \end{tabular} & 3.0$^{+0.7}_{-0.5}$ & 2.4$^{+0.5}_{-0.4}$ & 2.0$^{+0.4}_{-0.3}$ & 4.0$^{+0.9}_{-0.6}$ & 5.9$^{+1.1}_{-1.0}$ \\
 \hline
\begin{tabular}{c} Rad+El \\ $dN_g/dy$ ($\times$ 1000) \end{tabular} & 1.0$^{+0.3}_{-0.1}$ & 1.0$^{+0.2}_{-0.2}$ & 0.9$^{+0.2}_{-0.2}$ & 1.3$^{+0.2}_{-0.2}$ & 1.5$^{+0.3}_{-0.3}$ \\
 \hline
\end{tabular}
\caption{Comparison of the extracted $dN_g/dy$ for the work considered
  here compared to PHENIX 0-5\% most central $\pi^0$ data.  Note
  especially the nearly factor of 3 difference in the last three
  columns when using Rad Only (radiative energy loss only).  Due to
  the use of the collinear approximation all three of these values
  should be considered equally valid determinations of the medium
  density.  The assumed infinite precision for the elastic channel
  makes the convolved Rad+El extraction of $dN_g/dy$ suffer a smaller
  systematic uncertainty than from those studied in the radiative
  channel alone.  The first two columns demonstrate the limited
  influence of the radiated gluon mass in measuring the medium
  density.  See the text for more details. 
}
\label{tab:extract}
\end{table}

Once the qualitative picture is fixed, jet tomography \cite{Gyulassy:2001zv,Gyulassy:2001nm}, the
\emph{quantitative} determination of bulk properties of the QGP
through the study of the attenuation pattern of high momentum
particles, becomes possible.  Jet tomography requires both high
precision experimental data and a precise theoretical understanding of
partonic energy loss.  Recent statistical analyses by PHENIX
\cite{Adare:2008cg,Adare:2008qa} assuming infinite precision for
theoretical calculations that 
assume a weakly-coupled QGP suggest that data are now certain
enough to extract medium properties to within $\sim20\%$.   
Given that no current perturbative energy loss model
satisfactorily and simultaneously describes more than one \highpt observable \cite{Horowitz:2009pw,Nagle:2009wr},
such as the level of suppression and azimuthal anisotropy of the
\highpt gluons, light quarks, and heavy quarks as measured through the
light meson \cite{Adler:2006bv,Lin:2008zzi,Abelev:2009wx,Afanasiev:2009iv} and non-photonic electron nuclear modification factor \cite{Adare:2006nq,Abelev:2006db,Adare:2009ic,Wang:2008kha}, $R_{AA}(\eqnpt,\phi)$, and the away-side suppression of jet-triggered hadrons, $I_{AA}$ \cite{Abelev:2009gg}, it is, perhaps, premature to claim that even a qualitative
understanding of the jet quenching of the QGP exists.  Nevertheless,
continuing with the assumption that the current perturbative quantum chromodynamics (pQCD) models accurately
describe the physics, such quantitative extractions of medium
properties 
can only be meaningful when the systematic uncertainties in the
theoretical calculations and modeling are both known and included in
the systematic uncertainties in the extracted quantities.   
While there have been some previous qualitative estimates of the
theoretical uncertainty stemming from the running of the strong
coupling \cite{Horowitz:2006ya,Wicks:2007am}, the probability leakage
of the Poisson convolution \cite{Dainese:2004te,Eskola:2004cr}, and a
phenomenological IR cutoff imposed to approximate the effects of a
non-zero thermal gluon mass \cite{Baier:2001yt}, one of the main
purposes of this paper is to begin the process of rigorously
quantifying the combined experimental uncertainties and systematic
theoretical uncertainties in extracted medium parameters. 

In particular, we investigate the theoretical uncertainties due to the
collinear approximation and due to different assumptions regarding the
thermal mass of the radiated gluon in the opacity expansion approach
\cite{Gyulassy:1999zd,Gyulassy:2000er,Wiedemann:2000za,Djordjevic:2003zk,Djordjevic:2008iz}
for calculating the radiative energy loss of \highpt partons in
pQCD. We also asses how those uncertainties
change when elastic energy loss is included
\cite{Mustafa:2004dr,Wicks:2005gt}.  A summary of our results is shown in
\tab{tab:extract}.  Most strikingly, we find that keeping only the
lowest order term in collinearity results in $\sim200\%$ systematic
uncertainty on the value of the extracted gluon rapidity density for
central Au$\,+$Au collisions at top RHIC energies.  While this
paper does not quantify the uncertainties associated with the collinear  
approximation for other pQCD radiative energy loss models,
\cite{Gyulassy:1999zd,Dainese:2004te,Zhang:2007ja,Qin:2007rn}, 
the uncertainties in those models are almost certainly similar to
the results quoted here. In particular, the 
sizable discrepancies between the extracted medium properties by
different energy loss groups
\cite{Adare:2008cg,Adare:2008qa,Bass:2008rv} are probably within the
current theoretical systematic uncertainty.   

\begin{figure}[!t]
\centering
\includegraphics[width=\columnwidth]{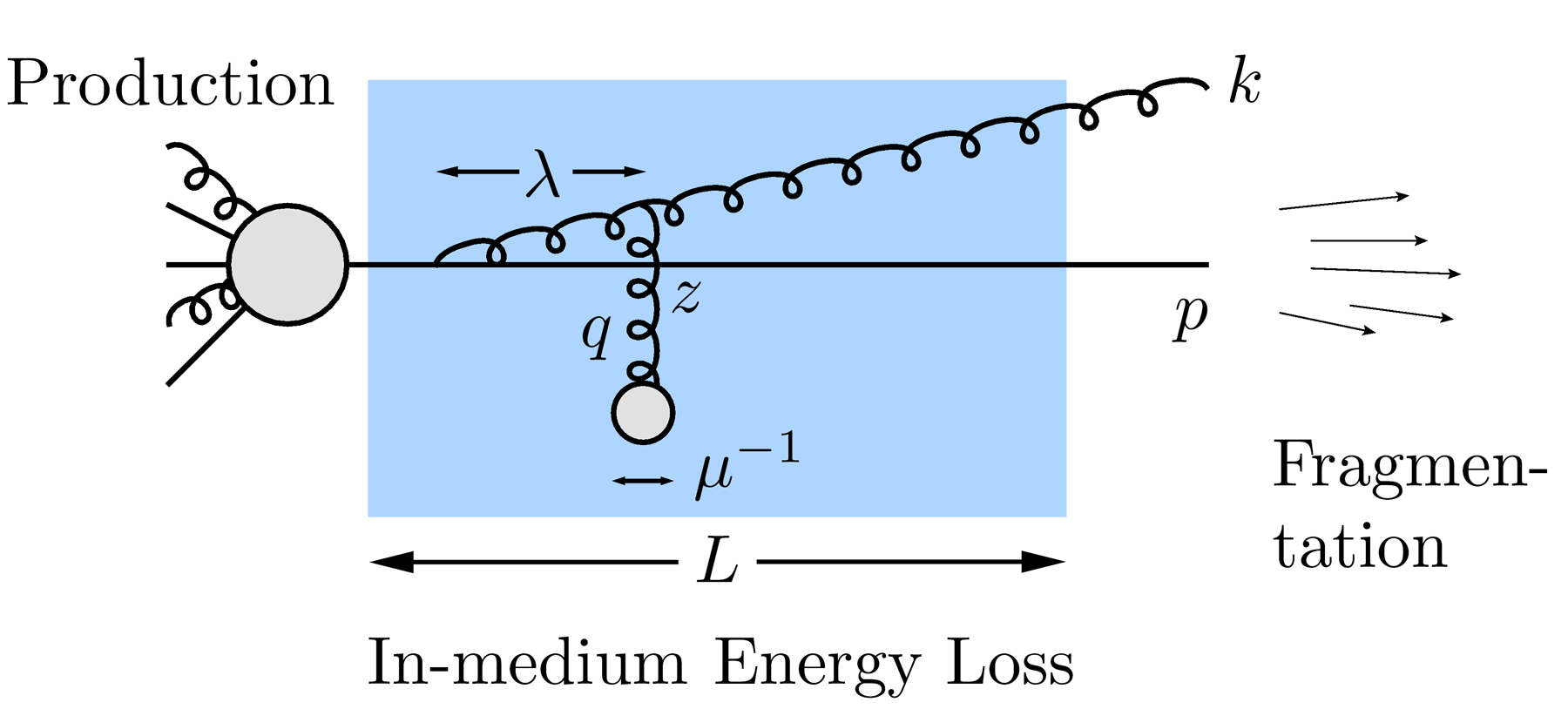}
\caption{\label{fig:diagram}
(Color online) Cartoon of the production, in-medium energy loss, and fragmentation processes that may occur perturbatively for a \highpt parton produced in a heavy ion collision.  The momentum labels are $p$ for the outgoing parent parton, $k$ for the medium-induced bremsstrahlung gluon, and $q$ for the momentum transfer between an in-medium soft degree of freedom and the \highpt parent parton.  Note the ordering of length scales displayed, $\mu^{-1}\ll\lambda\ll L$.
}
\end{figure}

There are, generally speaking, four main pQCD-based radiative energy
loss formalisms applied to URHIC: opacity expansion (GLV)
\cite{Gyulassy:1999zd,
Gyulassy:2000er,
Gyulassy:2001nm,
Djordjevic:2003zk,Djordjevic:2008iz}, multiple soft scattering
(BDMPS-Z-ASW)
\cite{Baier:1996kr,Baier:1996sk,Zakharov:1996fv,Zakharov:1997uu,Zakharov:1998sv,Wiedemann:2000ez,Wiedemann:2000za,Salgado:2002cd,Armesto:2003jh,Salgado:2003gb,Armesto:2004ud},
higher-twist (HT)
\cite{Guo:2000nz,Wang:2001ifa,Zhang:2003yn,Majumder:2004pt}, and thermal field theory (AMY)
\cite{Arnold:2000dr,Arnold:2001ba,Jeon:2003gi,Turbide:2005fk}.  In
this work we focus quantitatively on the opacity expansion approach;
see \fig{fig:diagram} for a cartoon of the physics involved and
visualizations of the key variables discussed in this text.  Within the GLV formalism
one derives an expression for the single inclusive radiated gluon
spectrum, $dN_g/dx$, that is folded into a Poisson convolution
\cite{Gyulassy:2001nm} for the distribution of total radiated energy
by the parent parton (see below).  Like all radiative energy loss
models, the formalism makes the eikonal approximation, namely that the
parent parton has a sufficiently high energy that its path is
approximately straight and that the interference from the
away-side jet, of $\mathcal{O}(1/E)$, may be safely ignored
\cite{Kovner:2003zj}. The formalism also neglects contributions from four-gluon vertices and 
assumes that the parent parton suffers independent, path-ordered
collisions.  
These choices are justified in the opacity expansion approach by the
assumption of a 
medium composed of Debye-screened scattering centers whose screening
length $\mu^{-1}\ll\lambda$ is much smaller than the mean free path of
the parent parton \cite{Baier:1996kr}. Such an ordering of scales
is consistent with thermal field theoretic estimates for $\mu$ and
$\lambda$ \cite{Wicks:2008ta}. 

Current evaluations of the diagrams resulting from the opacity
expansion formalism drop a significant number of terms that are assumed to be small.  These approximations were made for analytic simplicity,
but they do not seem to be inherently required.  Specifically, GLV takes:
(1) the collinear approximation, $k_T\ll xE$, where $k$ is the
momentum of the radiated gluon and $k_T$ its component transverse to
the motion of the parent parton; (2) the parent parton path length much
longer than the gluon mean free path, $L\gg\lambda$; and (3) the soft
radiation limit, $x\ll1$, where $x$ is the momentum fraction taken
away by the radiated gluon (discussed further below).  We note that
all these assumptions are also made in the BDMPS and AMY
formalisms\footnote{In GLV and BDMPS (2) is used to neglect poles from
  propagators multiplied by $\exp(-\mu\Delta
  z)\approx\exp(-\mu\lambda)\ll1$, where $\Delta z$ is the distance
  between successive scattering centers; this approach is probably invalid for
  $L\lessim\lambda\sim1$~fm.  On the other hand AMY uses the central
  limit theorem in its Langevin approach and corrections are likely
  for $L\lessim30\lambda\sim30$~fm; this extra long path length is
  also required by the neglect of the interference between vacuum and
  in-medium induced radiation.}.  HT makes a similar assumption to
(2), does not make the soft gluon approximation (3), but does assume
collinearity (1).  After discussing (1) in greater detail we will
briefly touch on (2) and (3) further below. 

This work focuses on the collinear approximation, $k_T\ll xE$, and,
more generally, the effects of limiting the phase space into which
gluon bremsstrahlung is allowed to radiate.  The current derivations using the opacity expansion formalism yield a single inclusive gluon radiation kernel, $dN_g/dxdk_T$, that knows nothing about the
approximations used in its derivation; in calculating $dN_g/dx$ the collinear approximation is enforced phenomenologically with a UV
cutoff in the $k_T$ integration.  We find that the $dN_g/dxdk_T$ kernel maximally
violates the assumption of collinearity at small $x$.  It is not surprising,
therefore, that $dN_g/dx$, and observables dependent on it such as $R_{AA}$, are
highly sensitive to the $\mathcal{O}(1)$ variations of the cutoff one
explores in estimating a systematic theoretical uncertainty.  Even
worse, as we discuss further below, the use of only the lowest order
collinear term means that collinearly equivalent definitions of $x$
yield values of medium density extracted from data that differ by $\sim100\%$.  

Previous work \cite{Baier:2001yt} that investigated a phenomenological
$k_T$ cutoff in the IR to approximate a thermal mass of the radiated
gluon found a similarly strong sensitivity to the specifics of the
cutoff.  However, we find that with an explicit derivation
of energy loss for non-zero gluon mass instead of a cutoff imposed by hand \emph{a
  posteriori}, surprising and non-trivial cancellations yield an
energy loss that has little sensitivity to the exact value of $m_g$. 
\section{Other Sources of Uncertainty}
Before returning to the primary focus of the paper---uncertainties
introduced by the collinear approximation---it is worth enumerating
here other potential sources of theoretical uncertainty that we do not
attempt to quantify in this work but that, nonetheless, could be quite
large.  

In this paper, we calculate radiative energy loss using the first order expression from the opacity expansion; i.e.\ the single inclusive gluon radiation spectrum is derived by scaling up diagrams with only a single in-medium scattering for the parent parton or its bremsstrahlung radiation, as is depicted in \fig{fig:diagram}, by the average number of scatterings, $L/\lambda$.  
The $n^\mathrm{th}$ order in opacity explicitly evaluates the interference terms, neglected in the lower order expressions, from diagrams with $n$ in-medium scatterings.  Within the set of approximations given above, and assuming the medium consists of Debye-screened static scattering centers, the GLV formalism yields a closed form expression for $dN_g/dx$ to all orders in opacity.  Numerical study of $dN_g/dx$ \cite{Gyulassy:2000er,Gyulassy:2001nm,Wicks:2008ta} suggests that it does receive corrections from higher orders but that these are relatively small---around the $\sim30\%$ level for relatively long paths of $L = 5$~fm and smaller for shorter paths---for RHIC- and LHC-like conditions.

The Poisson convolution \cite{Baier:1996kr,Gyulassy:2001nm} of the
$dN_g/dx$ kernel is an attempt to approximate the full probability
distribution of radiative energy loss, $P(\epsilon)$, where the
fraction of radiated energy is defined by $E_f =
(1-\epsilon)E_i$.  The convolution assumes independent, incoherent
emission of gluons; the effect of neglecting the interference between
two or more emitted gluons is not currently known.  There are also
uncertainties associated with ``probability leakage''
\cite{Gyulassy:2001nm,Dainese:2004te}: in model implementations the
parent parton energy is often not dynamically updated throughout the
convolution leading to the possible violation of energy conservation
(i.e.\ $P(\epsilon)$ has support
for $\epsilon>1$).  In this work, we assign the total weight 
from the convolution at $\epsilon > 1$ to a delta function centered at
$\epsilon = 1$ (complete stopping), the so-called 
``non--re-weighted'' approach \cite{Dainese:2004te}.  
Studies have shown that this procedure reproduces reasonably well distributions
obtained from Poisson convolution procedures that dynamically update the parent
parton energy \cite{Wicks:2008}.  Moreover, for energy loss
calculations based on the GLV formalism, the leakage tends to be small. 
When elastic energy loss is included this leakage is even smaller
\cite{Horowitz:2009}.   

A realistic model of jet quenching in URHIC requires a
weighted averaging over a large range of medium path lengths $0\lessim L\lessim 12~{\rm  fm}$.  QGP temperatures at RHIC energies lead to a
gluonic mean free path of $\lambda\sim1-2$~fm \cite{Wicks:2008ta}.  Additionally, even in the most
central collisions, a significant portion of path lengths have
$L\lessim2$~fm, and the bias towards the surface from energy loss
makes these shorter paths even more important \cite{Dainese:2004te,Wicks:2005gt}.  The assumption,
$L\gg\lambda$, is therefore violated for a large fraction of energy
loss inducing processes.  In the $R_{AA}$ calculations of this work, we
simply apply the GLV formalism to all path lengths.  There is an additional
uncertainty associated with mapping the realistic medium density of
heavy ion collisions---with its approximate Bjorken expansion and
non-trivial, time-dependent transverse density profile---into the
static, uniform ``brick'' problem in which the analytic formulae were
derived.

The small $x$ approximation seems reasonable at RHIC energies where
the single inclusive gluon radiation spectrum $dN_g/dx$ peaks at
$x\sim\mu/E\lessim0.05\ll1$.  The small-$x$ approximation should apply even
better at LHC, where $E$ will be larger by a factor $\sim10$.   
However, the Poisson convolution widens the original
$dN_g/dx$ spectrum---no matter how low the $x$ value of the peak of
the spectrum---thus distributing its weight out to larger values
$x$.  In this way the convolution introduces a sensitivity to the
large $x$ region of $dN_g/dx$ that is not well controlled due to the
$x\ll1$ assumption. 

Probably the largest source of uncertainty, likely bigger even than
that due to the collinear approximation that we detail in this paper,
will result from values chosen for and/or the running of \alphas \cite{Horowitz:2006ya,Wicks:2007am}; since radiative energy loss varies
as $\alpha_s^3$, it is highly sensitive to changes in the value of the
coupling.  One factor of $\alpha_s$ arises from the emission of the gluon
bremsstrahlung while the other two result from interacting with the soft
degrees of freedom in the medium; see \fig{fig:diagram}.  For these
last two couplings it is not even clear what scale should be chosen
for the running.  Nevertheless, the momentum transfer $q$ will always
probe soft scales on the order of $\mu\sim0.5$~GeV
$\gtrsim\Lambda_\mathrm{QCD}\sim0.2$~GeV, and non-perturbative physics
will be involved.  The lack of a proof of factorization, whereby these
non-perturbative contributions necessarily become small at high
energies, is of course not particularly comforting.
\section{Opacity Expansion Energy Loss Kernel and \texorpdfstring{$x$}{x}}\label{sec:kernels}
The first order in opacity result for the radiated gluon spectrum
assuming massive quarks and gluons and using
the Gyulassy-Wang model of static scattering centers
\cite{Gyulassy:1993hr}, to lowest order in $x$ and $k_T/xE$, where
$k_T=|\wv{k}|$, is given by \cite{Djordjevic:2003zk}
\be
\setlength{\jot}{6pt}
\begin{split}
\label{eq:DGLV}
x \frac{dN_g}{dx} & = \frac{C_R \alpha_s}{\pi}\frac{L}{\lambda}\int \frac{d^2\wq}{\pi} \frac{2d^2\wk}{\pi} \frac{\mu^2}{\big(\wq^2+\mu^2\big)^2} \\
& \times \frac{\wk\cdot\wq(\wk-\wq)^2-\beta^2\wq\cdot(\wk-\wq)}{\big[(\wk-\wq)^2+\beta^2\big]^2\big(\wk^2+\beta^2\big)} \\
& \times \int dz \left[ 1-\cos\left( \frac{(\wk-\wq)^2+\beta^2}{2Ex}z \right)\right] \rho(z).
\end{split}
\ee
Here, $\beta^2 = x^2 M^2 + m_g^2$, where $M$ is the mass of the parent
parton, and $m_g$ is the mass of the radiated gluon; $C_R=C_F$ for a parent quark, and $C_R=C_A$
for a parent gluon.  The corresponding result for massless parent
quarks/gluons and/or massless radiated gluons is obtained by setting
$M$ and/or $m_g$ to zero in \eq{eq:DGLV}, respectively.  After
temporarily taking the $q_T=|\wv{q}|$ medium exchange momentum limit
to infinity and making a change of variables (see
\cite{Djordjevic:2003zk} for more details) and additionally assuming an exponentially decaying distribution, $\rho(z)$, for the distance to the scattering center\footnote{In general it is necessary to specify
  the distribution in the \emph{difference} in distance between
  successive scattering centers.  In the first order in opacity case,
  with only one scattering center, this difference in distance is
  between the scattering center and the production point.  One can
  always take the production point to be at $z=0$, and thus in this case the
  specified distribution is the distance to the first scattering center.}, \eq{eq:DGLV} becomes  
\be
\begin{split}
\label{eq:dndx}
& \frac{dN_g}{dx} = \;  
\frac{8C_R\alpha_s\mu^2}{\pi\,x}\frac{L}{\lambda}\int dq_T dk_T \frac{q_T^3}{(4xE/L)^2+(q_T^2+\beta^2)^2}\\[5pt]
& \; \times \frac{k_T}{k_T^2+\beta^2} \frac{k_T^2(k_T^2+\mu^2-q_T^2)+\beta^2(q_T^2+\mu^2-k_T^2)}{\big((k_T-q_T)^2+\mu^2\big)^{3/2}\big((k_T+q_T)^2+\mu^2\big)^{3/2}}.
\end{split}
\ee
We note that supposing the scattering center distribution to be
uniform instead of exponentially decaying appears to make only a small
difference in the gluon spectrum \cite{Horowitz:2009}.   

The integrand in \eq{eq:dndx} is both IR and UV safe.  In principle, we
could set the lower and upper $q_T$ and $k_T$ limits of integration to
0 and $\infinity$, respectively.  If the integrand were exact, then it
would have support only in the physical regions of the $q_T$ and $k_T$
integration space.  However, due to the small $x$ and collinear
approximations, the integrand violates kinematic limits: one can clearly see that the
integrand in \eq{eq:dndx} has support over all $q_T$ and $k_T$ (this
is also true of the unshifted integrand, \eq{eq:DGLV}).  We enforce
physicality in the hope of better approximating the exact result by
restricting the $q_T$ and $k_T$ integration region with cutoffs.  A
new result from this paper is a quantitative estimate of the systematic
error that results from allowing $\mathcal{O}(1)$ variations of these
cutoffs on  some observables and extracted medium parameters.  Surprisingly, we find that $dN_g/dx$ is highly sensitive to variations
in the $k_T$ UV cutoff. 

\begin{figure}[!t]
\centering
\includegraphics[width=.99\columnwidth]{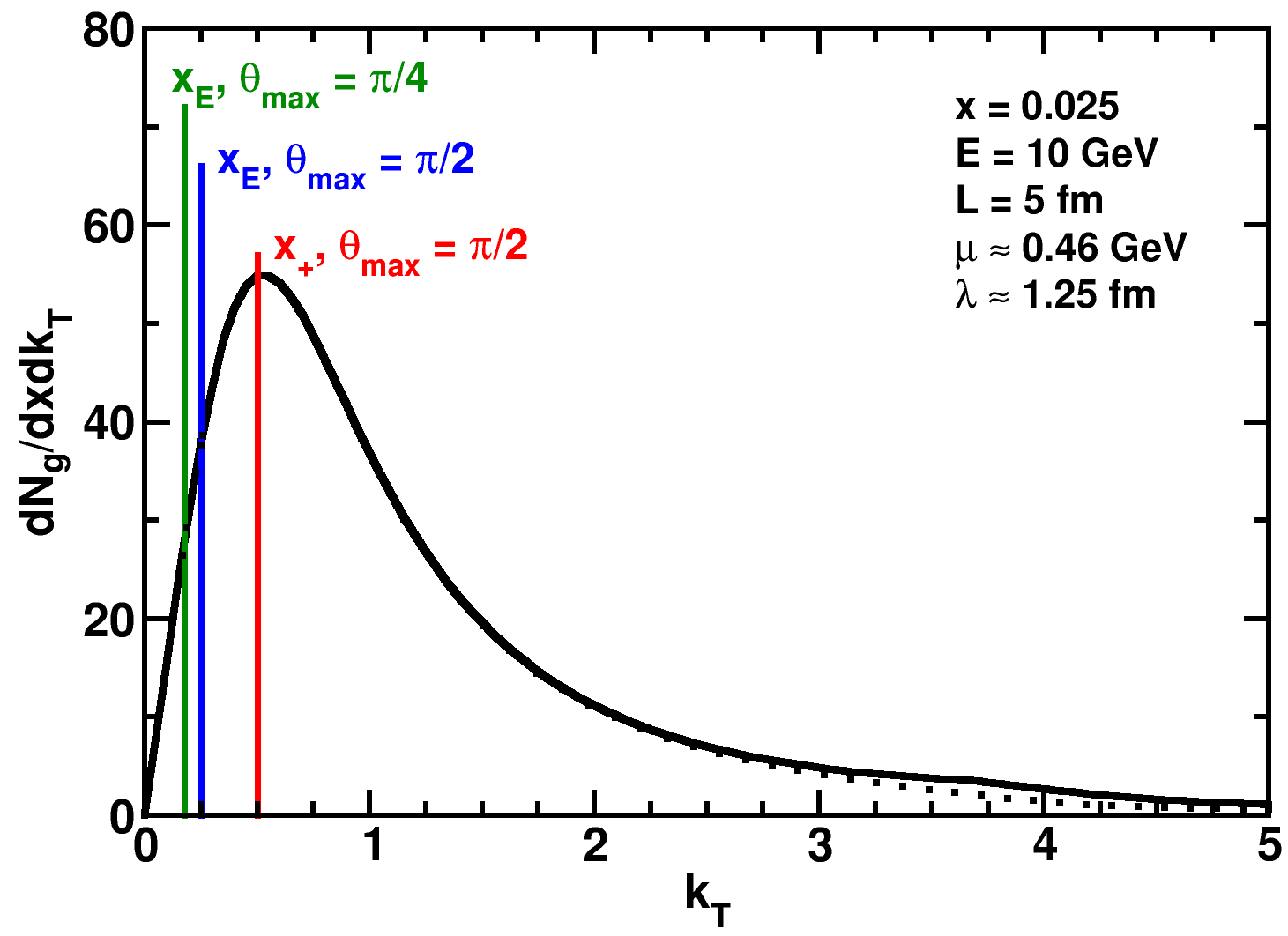}
\caption{\label{fig:dndxdk}
Plot of $dN_g/dxdk_T$ from \eq{eq:dndx} (solid curve) for a light
quark with all masses set to 0, $E=10$~GeV, $L=5$~fm, and 
representative values of $\mu\simeq0.46$~GeV and
$\lambda\simeq1.25$~fm for a medium density of $dN_g/dy = 1000$
similar to RHIC conditions 
\cite{Wicks:2005gt}.  
Vertical lines depict the values of $k_T$ used as cutoffs to enforce
collinearity in \eq{eq:dndx}.  
Note that with \mbox{$x = 0.025 \sim \mu/E$},
$dN_g/dxdk_T$ is large near $k_T\sim k_\mathrm{max}$, 
completely in contradiction with the collinear approximation.  
Dotted curve, from the unshifted integrand of \eq{eq:DGLV}, differs only slightly from the spectrum obtained from the shifted integrand of \eq{eq:dndx} (solid curve).
} 
\end{figure}

\begin{figure}[!htb]
\centering
\includegraphics[width=.99\columnwidth]{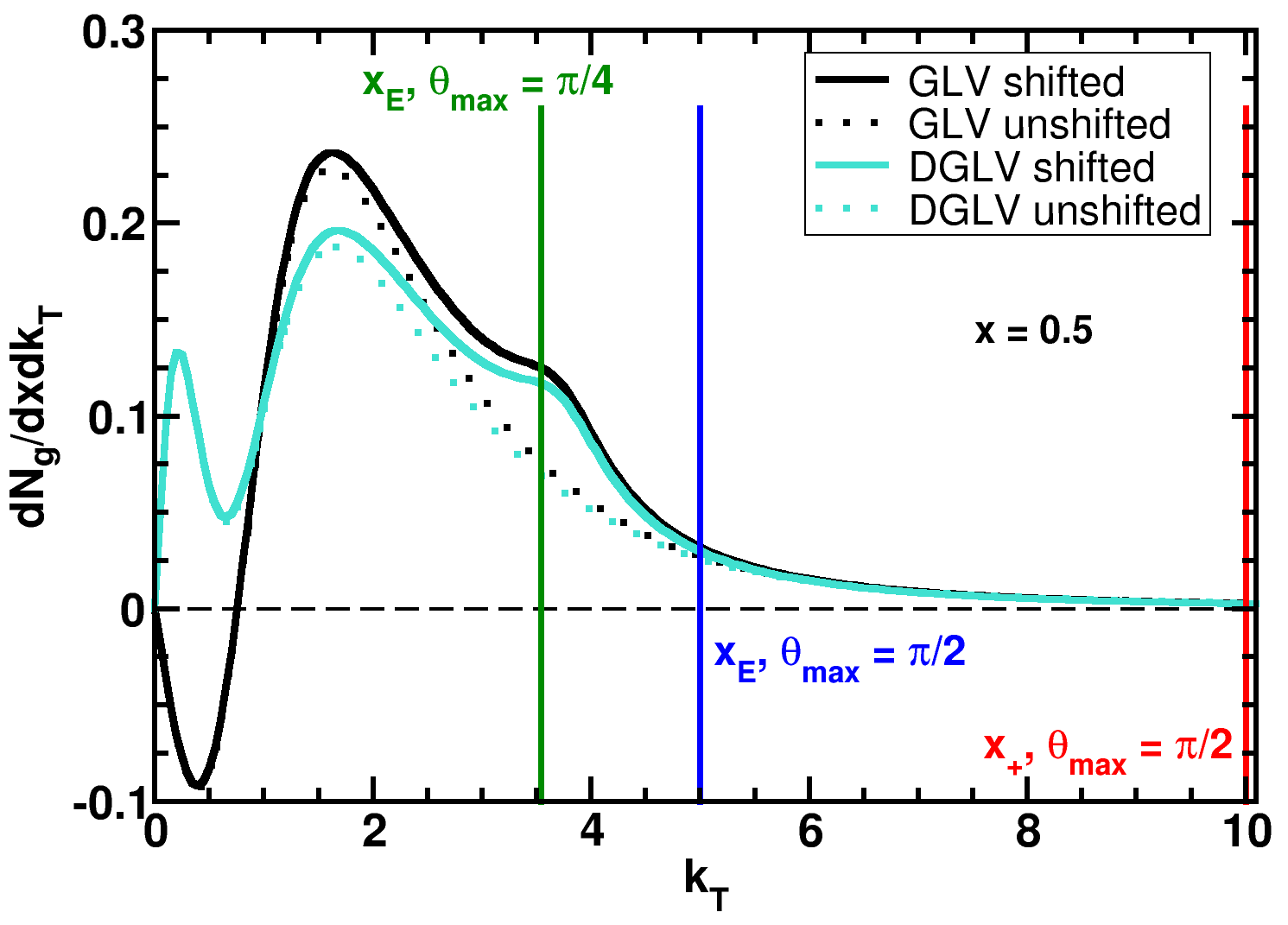}
\caption{\label{fig:dndxdktwo}
Comparison of  $dN_g/dxdk_T$ spectra as a function of $k_T$ with
$x=0.5$ between GLV  (all masses set to zero) and DGLV
(non-zero radiated gluon mass) formulations using both the shifted integrand of
\eq{eq:dndx} (solid curves) and the unshifted integrand of
\eq{eq:DGLV} (dotted curves). The horizontal dashed line at 0 is
meant to guide the eye.
The three vertical lines represent the choices of $k_\mathrm{max}$ described in the text.  $dN_g/dxdk_T$ better respects the collinear approximation for larger values of $x$ as it has little weight near $k_T\sim k_\mathrm{max}$. 
At small $k_T$ the radiated gluon mass, surprisingly, \emph{enhances} radiation for DGLV due to coherence effects; at larger $k_T$ the mass has the expected effect of suppressing $dN_g/dxdk_T$.
}
\end{figure}

For the $q_T$ integration we will make the usual choices $q_\mathrm{min}=0$
and $q_\mathrm{max} = \sqrt{3\mu E}$.  The value for $q_\mathrm{max}$
is, in principle, the maximum momentum transfer allowed using
relativistic kinematics, though it has been pointed out that this is
no longer correct with the change of variables
$\wv{q}\rightarrow\wv{q}+\wv{k}$ used in the final steps leading to
\eq{eq:dndx}\footnote{We thank Ulrich Heinz for finding 
  this error in logic.}.  Nevertheless, \eq{eq:dndx} is rather insensitive to the $q_T$ cutoff \cite{Djordjevic:2003zk}, and we show below explicit calculations of $dN_g/dxdk_T$ using the shifted and unshifted integrands of Eqs.~(\ref{eq:dndx}) and (\ref{eq:DGLV}), respectively.  These comparisons in Figs.~\ref{fig:dndxdk} and
\ref{fig:dndxdktwo} demonstrate that the shifted integrand with the unshifted $q_\mathrm{max}$ well reproduces the results of the unshifted integrand, albeit with minor artifacts.

Previous work by BDMS \cite{Baier:2001yt} found a strong sensitivity
to a phenomenological IR cutoff in the $k_T$ integration of the BDMPS
multiple soft scattering calculation, a residual of the original
vacuum radiation IR divergence. They also showed that the sensitivity
of a quenching factor to variations in this cutoff decreased with
increasing parent parton energy.  The IR cutoff was imposed in order
to approximate the influence of a non-zero thermal gluon mass; we find
that including a gluon mass at the level of the gluon propagators \cite{Djordjevic:2003zk} decreases the
sensitivity of the energy loss and calculated $R_{AA}$ (see
\tab{tab:extract}) to the choice of gluon mass.  This result is due to
a non-trivial cancellation of effects that we will discuss further below.   

We do find, however, that the radiated gluon spectrum and resulting
parton energy loss are extremely sensitive to the choice of the UV
cutoff in the $k_T$ integration.  Furthermore we find a sensitivity to the particular interpretation of $x$ used in relating the components of $k$ to the components of $p$.  
It turns out that the derivations of $dN_g/dx$ in the literature have used two different, albeit equal to
lowest order in collinearity, definitions of $x$.  
We will use the light-cone normalization $p^\pm=p^0\pm p^z$, with inverse
$p^{0,\,z}=(p^+\pm p^-)/2$, and denote $p^0=E$ and $p^+=E^+$.  With
these conventions the two definitions of $x$ are: (1) the 
fraction of plus momentum carried away by the gluon, $x = x_+ \equiv
k^+/E^+$ \cite{Gyulassy:1999zd,Gyulassy:2000er}; and (2) the fraction
of energy carried away by the gluon, $x = x_E \equiv k^0/E$
\cite{Wiedemann:2000za}.  In the usual notation where parentheses
designate four-momenta, $k = (k^0,\,k^z,\,\wv{k})$, square brackets
light-cone momenta, $k=[k^+,\,k^-,\,\wv{k}]$, and bold-faced variables
are the transverse 2-vectors ($k_T = |\wv{k}|$), we find that the
radiated massless gluon has on-shell momentum 
\be
\label{eq:k}
k = (x_E E,\, \sqrt{(x_E E)^2-\wv{k}^2},\, \wv{k}) = [x_+ E^+,\, \frac{\wv{k}^2}{x_+ E^+},\, \wv{k}].
\ee
With these definitions of $x_+$ and $x_E$ one may derive the exact relationships
\begin{align}
x_+ & = \frac{1}{2}\,x_E\left(1+\sqrt{1-\left(\frac{k_T}{x_EE}\right)^2}\,\right), \\[5pt]
x_E & = x_+\left(1+\left(\frac{k_T}{x_+E^+}\right)^2\right),
\end{align}
where we have assumed that the initial parent parton momentum is $P =
(E,\, E,\, \wv{0})$, and, hence, $E^+ = 2E$.  From these formulas, it is
easy to see that in the the collinear limit, $x_+=x_E$, so
it is natural that the integrand in \eq{eq:dndx} is the same
when derived using these two different definitions of $x$: in one case
terms of order 
$k_T/x_EE$ are dropped whereas in the other terms of order
$k_T/x_+E^+$ are dropped.  Thus $dN_g/dx_+dk_Tdq_T(x_+) =
dN_g/dx_Edk_Tdq_T(x_E)$ in the collinear limit.   

Two justifications for and derivations of a $k_T$ cutoff exist in the
literature: (1) when interpreting $x$ as $x_E$, $\kmax=x_EE$ keeps $k$
always real (note that in the $x=x_+$ representation $k$ is real
regardless of the value of $k_T$) \cite{Armesto:2003jh}; (2) when
interpreting $x$ as $x_+$, $\kmax=x_+E^+$ enforces forward emission
($k^+\ge k^-$) \cite{Gyulassy:2000fs}\footnote{An even more
  restrictive  cutoff for $k_T$ results if one also requires forward
  propagation of the 
  parent parton \cite{Gyulassy:2001nm}.  Surprisingly this tighter
  cutoff, which forbids support for $dN/dx$ for $x>1$ and therefore
  enforces energy conservation, leads to only a small change in
  $dN_g/dx$ \cite{Horowitz:2009}, and we will not use it here.}.  
In this work, we will interpret $k_\mathrm{max}$ as enforcing
consistency with the collinear approximation.  It is, then, useful to
determine that cutoff from a physical condition independent of the
interpretation of $x$; we choose to set $k_\mathrm{max}$ by requiring
the emitted radiation be within a cone of angle $\theta_\mathrm{max}$
centered on the direction of the parent parton. In this way we may
control the collinearity of the radiation by varying the maximum angle
of emission allowed.  
Simple trigonometry relates 
\be
k_\mathrm{max} = \left\{ \begin{array}{ll}
x_+ E^+ \tan (\theta_\mathrm{max}/2),\quad & x=x_+, \\
x_E E \sin (\theta_\mathrm{max}),\quad & x=x_E.
\end{array}\right.
\ee
In particular, when $\theta_\mathrm{max}=\pi/2$
\be
\label{eq:kmaxpibytwo}
k_\mathrm{max} = \left\{ \begin{array}{ll}
x_+ E^+ \, = \, 2 x_+ E,\quad & x=x_+, \\
x_E E,\quad & x=x_E,
\end{array}\right.
\ee
and we see that the two original justifications of the
$k_\mathrm{max}$ cutoff are manifestations of the same physical
condition, namely forward emission.  It is worth emphasizing that with $k_\mathrm{max}$ given by
\eq{eq:kmaxpibytwo}, for any numerically equal
value of $x_+$ and $x_E$ the integrand in \eq{eq:dndx} is integrated out
twice as far in $k_T$ when interpreting $x$ as $x_+$ as opposed to
interpreting $x$ as $x_E$.  We note that a previous study
\cite{Salgado:2003gb} also varied the upper limit of the $k_T$
integration.  However that work investigated the radiation into jet
cones of various sizes; here we are interested in quantifying the
impact of regions of $k_T$ space over which there is currently little
or no theoretical control. 

\begin{table}[!t]
\begin{tabular}{|c|c|c|c|c|}
 \hline
Name & $x$ & $\phantom{\;}$Jacobian$\phantom{\;}$ & $\phantom{\;}\theta_\mathrm{max}\phantom{\;}$ & $m_g$ \\
\hline\hline
$x_+$, $m_g=\mu/\surd2$ & $\phantom{\;}x_+\phantom{\;}$ & No & $\pi/2$ & $\phantom{\;}\mu/\surd2\phantom{\;}$ \\
\hline
$x_+$, $m_g=0$ & $x_+$ & No & $\pi/2$ & $0$ \\
\hline
$\phantom{\;}x_+(x_E)$, $\theta_\mathrm{max}=\pi/2\phantom{\;}$ & $x_+$ & Yes & $\pi/2$ & $0$ \\
\hline
$x_E$, $\theta_\mathrm{max}=\pi/2$ & $x_E$ & No & $\pi/2$ & $0$ \\
\hline
$x_E$, $\theta_\mathrm{max}=\pi/4$ & $x_E$ & No & $\phantom{\;}\pi/4\phantom{\;}$ & $0$ \\
 \hline
\end{tabular}
\caption{Descriptions of the five main radiative energy loss implementations investigated in this work and their names when listed in diagrams.  See the text for details.}
\label{tab:models}
\end{table}

If the integrand of \eq{eq:dndx} respected the assumptions that went
into its derivation over the entire integration region, then it would 
have little weight for $k_T\sim xE$.  
However, \fig{fig:dndxdk} shows that there are values of $x$ for which $dN_g/dxdk_T$ has large contributions from $k_T\sim xE$.  In fact, we should expect such a result: the Debye mass  
$\mu$ is the natural scale for $k_T$, and $dN_g/dxdk_T$ will always
have a significant weight at $k_T\sim\mu$. As a result, varying
$\kmax\sim xE$ will always lead to large changes in $dN_g/dx$
near $x\sim\mu/E$.  In the kinematic regime relevant for RHIC and LHC,
numerical study suggests that (1) $dN_g/dx(x)$ reaches its maximum at
$x\sim\mu/E$ and (2) $dN_g/dxdk_T(x\sim\mu/E,\, k_T)$ reaches its
maximum at $k_T\sim k_\mathrm{max}$; for $x\sim\mu/E$ the maximum
value of $dN_g/dx$ rises quadratically with $k_\mathrm{max}$.
These observations imply a dramatic sensitivity of $dN_g/dx$---and any quantities derived from it---to the precise choice of
$k_\mathrm{max}$.  Increasing the radiating parton energy decreases the region of $x$ over which $dN_g/dx$ is highly sensitive to the cutoff, $x\lesssim\mu/E$.  Naively, then, one might expect that increasing $E$ would decrease the sensitivity of observables calculated from $dN_g/dx$ to variations of $k_\mathrm{max}$.  
However, we will show below that even this expectation does not generally hold.

\fig{fig:dndxdktwo} shows that, as expected, as $x$ increases the assumption of collinearity, $k_T\ll xE$, becomes a better approximation: $dN_g/dxdk_T$ has little weight near $k_T\sim k_\mathrm{max}$, and, therefore, for these larger values of $x$ $dN_g/dx$ has less sensitivity to the exact choice of $k_\mathrm{max}$.  One may also observe in the figure the negative values of $dN_g/dxdk_T$ for the massless case at small values of $k_T$.  These negative values are due to the destructive interference between the $0^\mathrm{th}$ order vacuum production radiation (the QCD analog of the usual beta-decay radiation spectrum from QED) and the $1^\mathrm{st}$ order medium-induced radiation.  The interference is controlled by the ratio of the radiating parton path length $L$ and the coherence length,
\be
\tau_\mathrm{coh} = \frac{2xE}{q_T^2+\beta^2}
\ee
found by examining the cosine term in \eq{eq:dndx} ($q_T$ enters $\tau_\mathrm{coh}$ here due to the shift in integration variables $q_T\rightarrow q_T+k_T$ in going from \eq{eq:DGLV} to \eq{eq:dndx}).  For larger values of $x$ the coherence length, $\tau_\mathrm{coh}$ is longer, and the destructive interference more important; hence $dN_g/dxdk_T$ is negative for small values of $k_T$ for the massless results shown in \fig{fig:dndxdktwo} but not in \fig{fig:dndxdk}.  Including a non-zero mass for the radiated gluon reduces $\tau_\mathrm{coh}$ and therefore the influence of the destructive interference.  As a result, for small values of $k_T$ the $dN_g/dxdk_T$ spectrum is actually \emph{enhanced} for the massive as compared to massless case.  At larger values of $k_T$ the non-zero thermal gluon mass exhibits the expected effect of suppressing bremsstrahlung radiation.  Unfortunately, then, introducing a phenomenological $k_T$ cutoff in the IR into a massless $dN_g/dx$ formula does not capture well the complicated dynamics resulting from allowing the radiated gluons to pick up a non-zero thermal mass from the medium.  
More readily visible in \fig{fig:dndxdktwo} than in \fig{fig:dndxdk} is the removal of the minor artifacts from the shift in integration variable when evaluating the unshifted integrand of \eq{eq:DGLV}.  

The ambiguity in the interpretation of $x$ has consequences beyond the 
differences in the corresponding limits of the $k_T$ integration. The
gluon $dN_g/dx$ distributions obtained from Eq.~\ref{eq:dndx} using the two
different $x$ definitions imply different gluon energy spectra
and, thus, different amounts of jet quenching. These differences will clearly 
contribute to the systematic uncertainty in the interpretation of
experimental data using the opacity expansion formalism and  
any other energy loss formalism that invokes the same or similar
collinear approximations. 
Furthermore, in order to better understand the
consequences of the different physical pictures used in different
formalisms---e.g.\ by comparing the gluon spectra produced by GLV
(traditionally interpreting $x$ as $x_+$) and BDMPS-Z-ASW ($x$ as
$x_E$)---one wants all results in terms of the same variables.  As a first step we will examine an ``apples-to-apples'' comparison
between results derived from the two interpretations of $x$ solely within the opacity expansion approach. 

There is an additional problem with energy loss calculations that
interpret $x$ in \eq{eq:dndx} as $x_+$: evaluation of the Poisson
convolution $P(\epsilon)$, where $\epsilon$ is explicitly an
\emph{energy} fraction.  Previous models that assumed the $x_+$
interpretation in $dN_g/dx$ invoked the collinear approximation in
order to directly use $dN_g/dx_+(x_+)$ unmodified as the input for
the Poisson convolution.  As we have shown this is a poor assumption.   

In order to compare the two interpretations of $x$ within GLV and to quantify
the effect of assuming $dN_g/dx_+\approx dN_g/dx_E$ in finding
$P(\epsilon)$ we need to transform $dN_g/dx_+(x_+)$ to $dN_g/dx_E(x_E)$,
which requires the use of a Jacobian: 
\begin{align}
\label{eq:dndxpxe}
\frac{dN^J_g}{dx_E}(x_E) & \equiv \displaystyle{\int^{x_E E\sin(\theta_\mathrm{max})} dk_T \frac{dx_+}{dx_E}\frac{dN_g}{dx_+dk_T}}\big(x_+(x_E)\big),\\[-5pt]
\frac{dx_+}{dx_E} & = \frac{1}{2}\left[1+\left(1-\Big(\frac{k_T}{x_EE}\Big)^2\right)^{-1}\right].\label{eq:jacobian}
\end{align}
Note the change in the upper limit of integration in \eq{eq:dndxpxe}
as dictated by the basic rules of calculus.  The Jacobian,
\eq{eq:jacobian}, is strictly greater than one and is singular as
$k_T\rightarrow x_EE$; the competing effects between it and the
numerically smaller $k_\mathrm{max}$ on $dN^J_g/dx_E$ and energy loss
will be discussed in further detail below.   

\begin{figure}[!t]
\centering
$\begin{array}{l}
\includegraphics[width=\columnwidth]{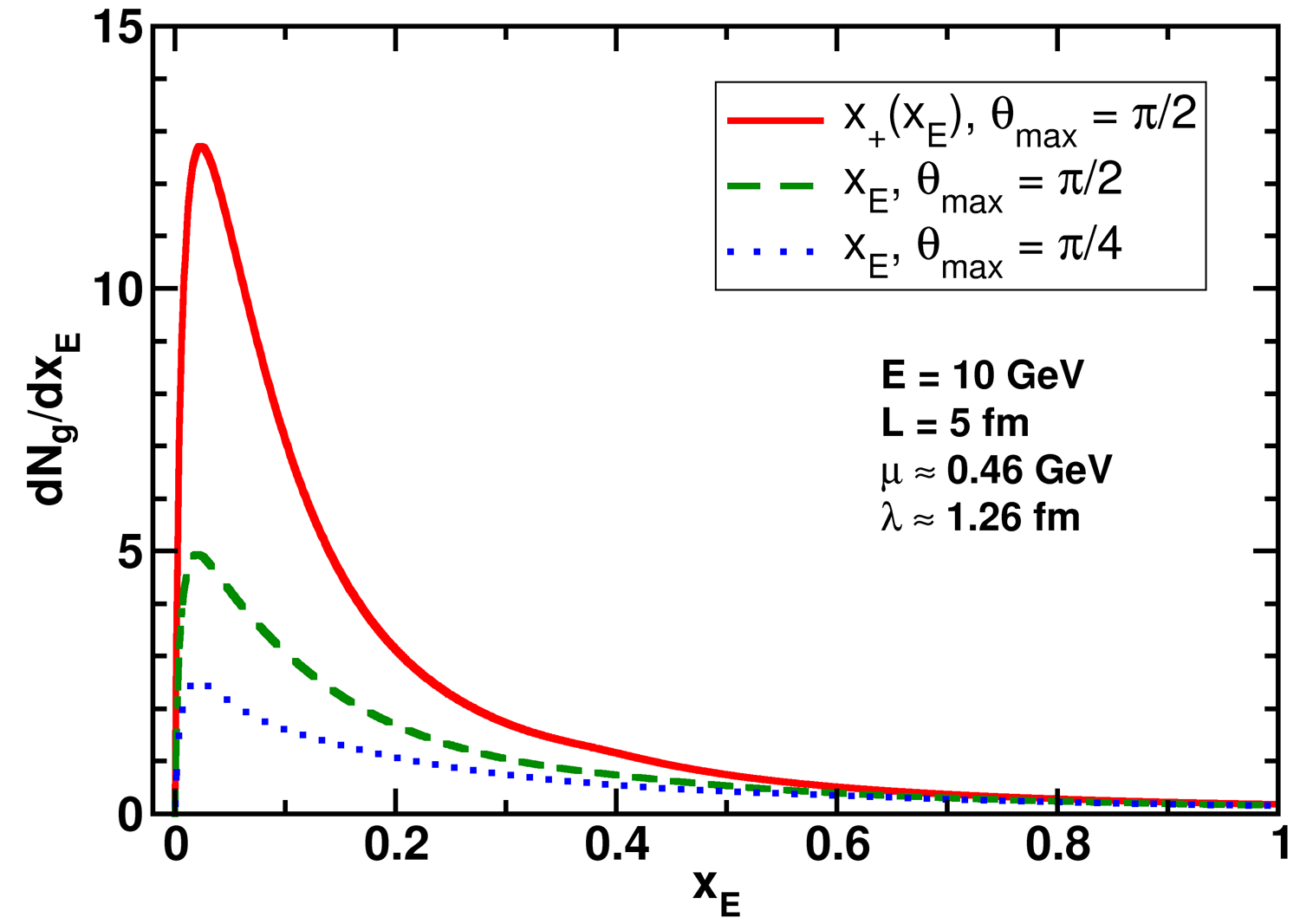} \\
\includegraphics[width=\columnwidth]{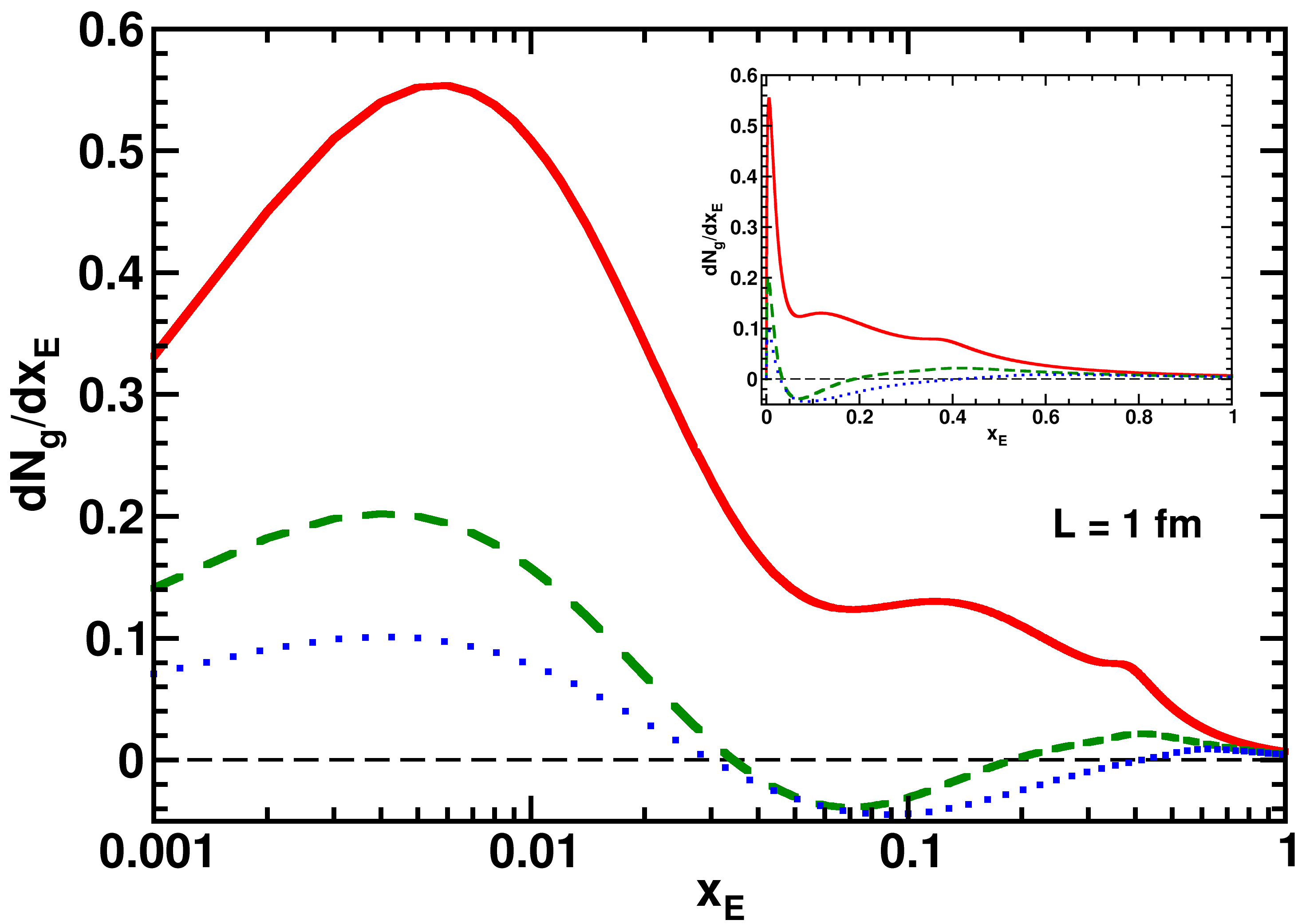}
\end{array}$
\caption{\label{fig:dndxe}
Comparison of $dN_g/dx_E(x_E)$ for a 10~GeV quark using typical RHIC medium parameters and \mbox{$L=5$~fm} (top), \mbox{$L=1$~fm} (bottom) for the collinearly equivalent expressions
$dN_g/dx$ from \eq{eq:dndx} with the $x=x_E$ interpretation and $x_+(x_E)$ from \eq{eq:dndxpxe}.
Also shown is $dN_g/dx$ from \eq{eq:dndx} with $x=x_E$ and
$\theta_\mathrm{max}=\pi/4$ in order to explore the systematic uncertainty in the choice of $k_\mathrm{max}$.  Inset shows results on a linear scale for clarity.
}
\end{figure}

\section{Quantitative Comparisons}

We now wish to quantify the effects of the different 
$x$ interpretations, $k_\mathrm{max}$ values, and gluon mass treatments on $dN_g/dx$ and its derived quantities, such as $P(\epsilon)$, the medium parameter $dN_g/dy$ extracted from $R_{AA}$, and the total radiated gluon multiplicity, $\langle N_g \rangle$.  We
will do so using five implementations of radiative energy loss; see
\tab{tab:models}.  In the
first three calculations we use the $x=x_+$ interpretation and
$\thetamax=\pi/2$; 
in the first two we do not include the effect of the $dx_+/dx_E$
Jacobian.  In the first calculation we assume radiated gluons acquire
a thermal mass on the order of the Debye scale, $m_g=\mu/\surd2$.  The
second calculation is
the same as the first but with $m_g\rightarrow0$.  The third is the
same as the second but includes the Jacobian transformation from $x_+$ to $x_E$.  In the last two
calculations we adopt the $x=x_E$ interpretation from the start: in the
fourth we take $\thetamax=\pi/2$ and the fifth $\thetamax=\pi/4$.   

\begin{figure}[!t]
\centering
$\begin{array}{l}
\includegraphics[width=\columnwidth]{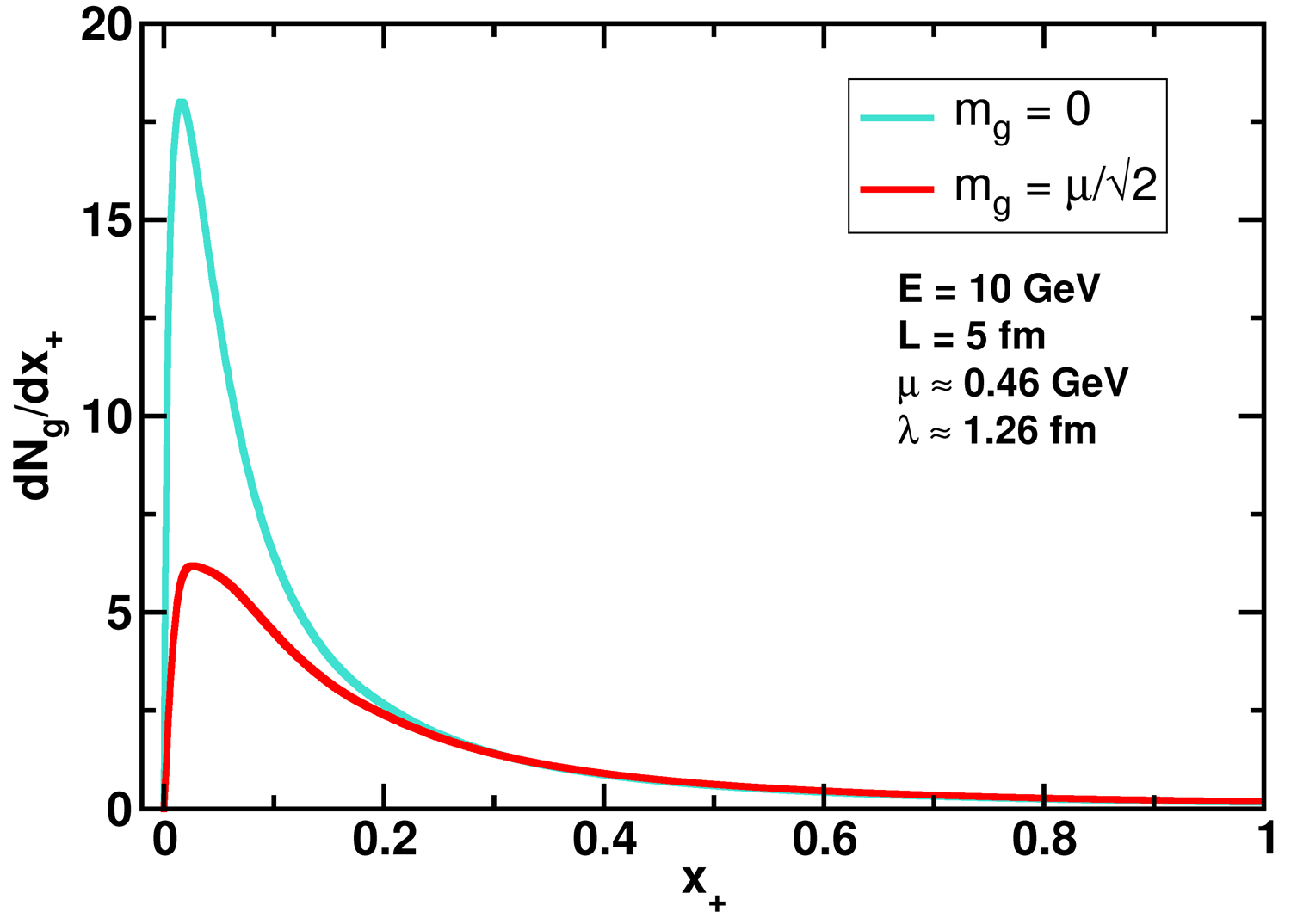} \\
\includegraphics[width=\columnwidth]{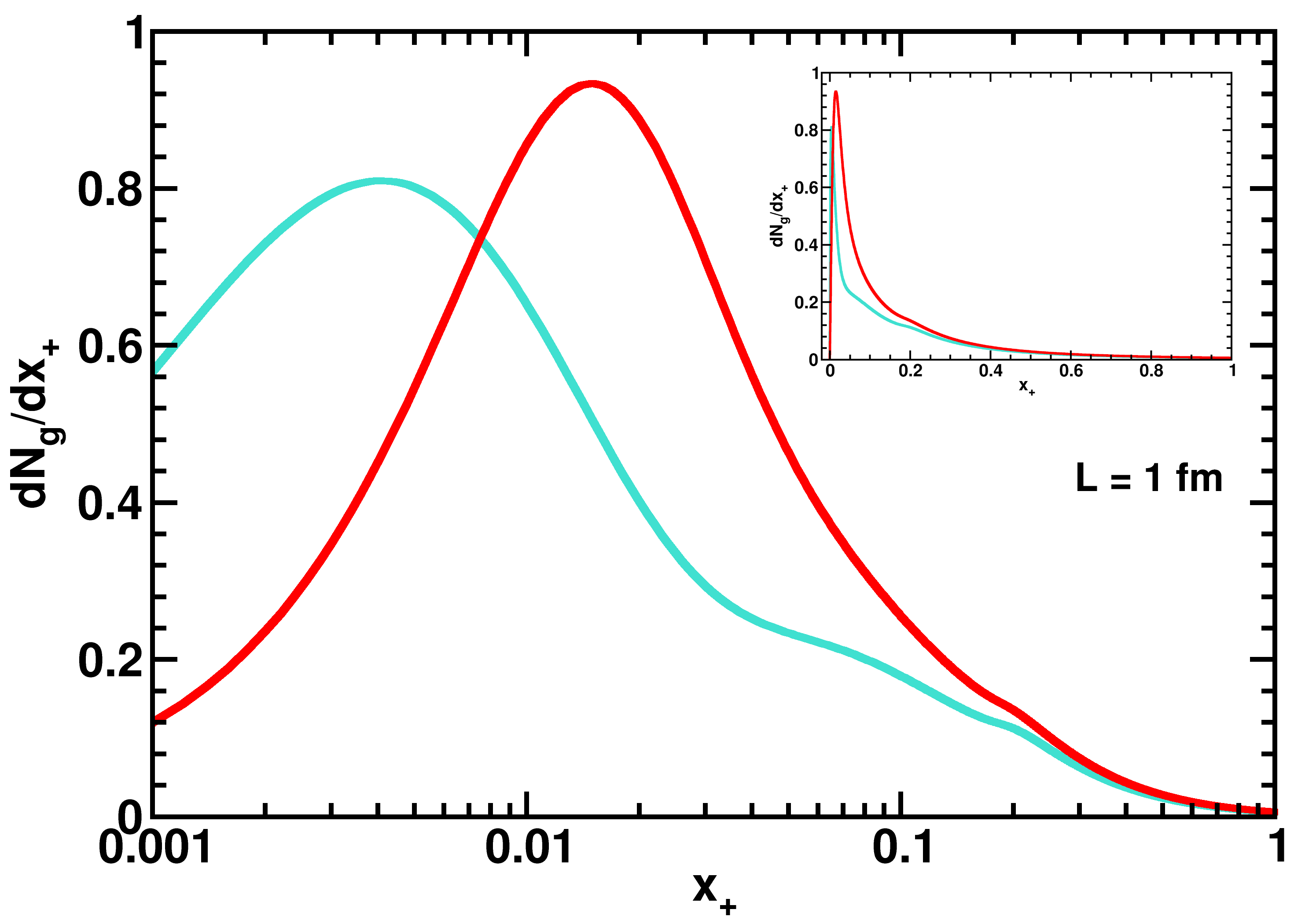}
\end{array}$
\caption{\label{fig:dndxmg}
Comparison of $dN_g/dx_E(x_E)$ from GLV and DGLV ($m_g = \mu/\surd2$) 
for a \mbox{$E=10$~GeV} quark using typical RHIC medium parameters and \mbox{$L=5$~fm} (top), \mbox{$L=1$~fm} (bottom).  Inset shows results on a linear scale for clarity.
Note the \emph{enhancement} of $dN_g/dx$ at \mbox{$L=1$~fm} for the massive case.
}
\end{figure}

We show in \fig{fig:dndxe} top (bottom) the dramatic difference between the
``equivalent'' $dN_g/dx_E$ spectra obtained using the two
interpretations of $x$ and taking $E=10$~GeV, $L=5$~fm ($L=1$~fm), $\mu\approx0.46$~GeV and 
$\lambda\approx1.26$~fm as representative RHIC values.  Also
shown in the figure is the large reduction in $dN_g/dx$ as
$\theta_\mathrm{max}$ is reduced from $\pi/2$ to $\pi/4$. That
reduction in the spectrum demonstrates the importance 
of large angle gluon emission despite the assumption of collinearity.
\fig{fig:dndxmg} (top) shows the dramatic reduction in $dN_g/dx$ with
the introduction of a thermal gluon mass for the above parameter
values; the bottom panel in the same figure shows that for shorter
path lengths, such as $L=1$~fm, the inclusion of a thermal mass for the
radiated gluon can actually enhance the emission of bremsstrahlung 
radiation.  

\begin{figure}[!thb]
\centering
\includegraphics[width=\columnwidth]{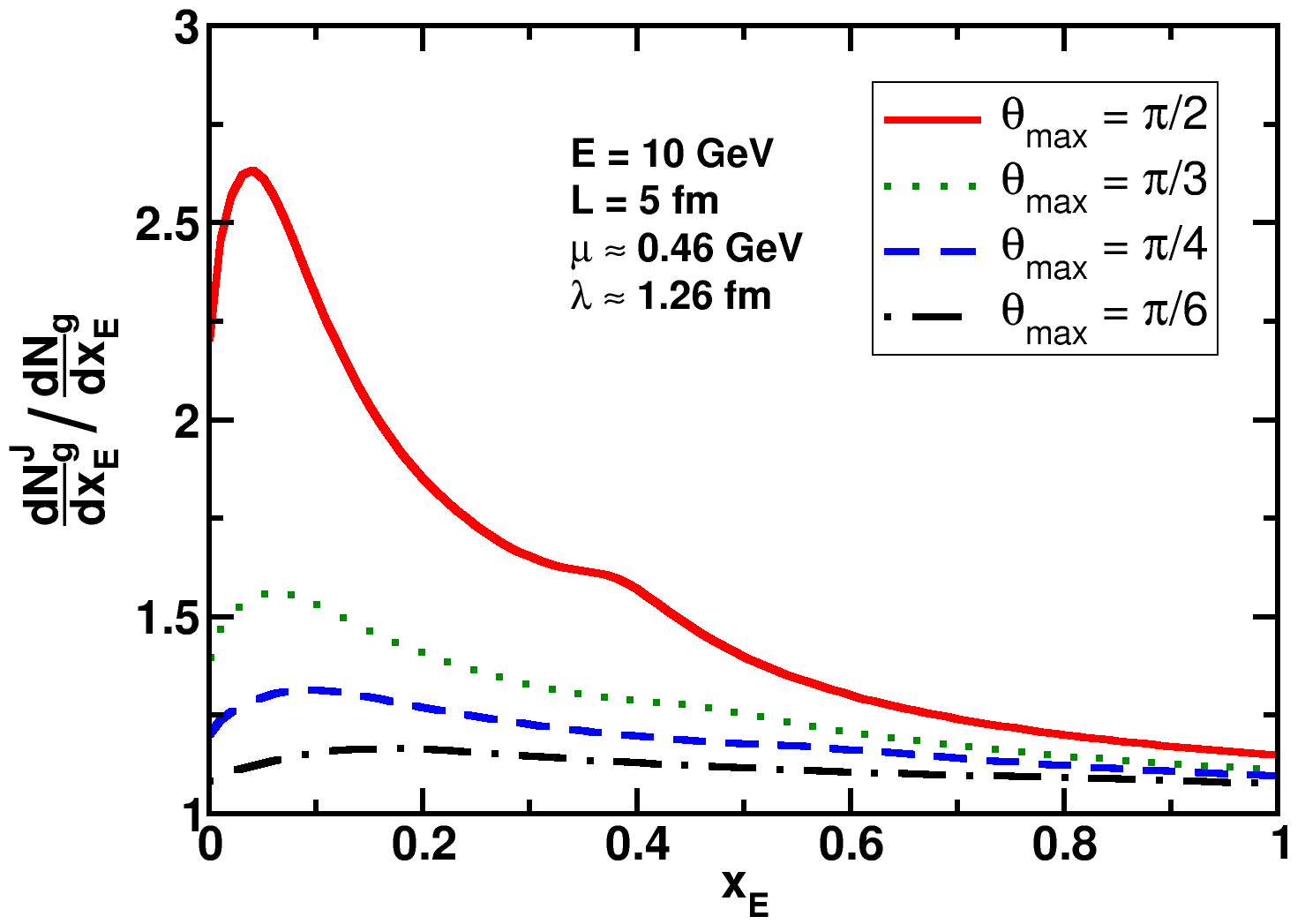}
\caption{\label{fig:dnrat}
Ratios of gluon spectra obtained from \eq{eq:dndxpxe}, $dN_g^J/dx_E(x_E)$, to that
obtained from \eq{eq:dndx} with $x=x_E$, $dN_g/dx_E(x_E)$, as a function of $x_E$ for different choices of $\theta_\mathrm{max}$.  
Smaller maximum opening angles produce a decreased sensitivity to the collinear approximation. 
}
\end{figure}

The only difference between the $x_E$ and $x_+(x_E)$ curves,
theoretically indistinguishable to lowest order in collinearity, in
\fig{fig:dndxe} comes from terms of
$\mathcal{O}\big((k_T/x_EE)^2\big)$ and higher.  \fig{fig:dnrat}
quantifies the reduction of the effect of these terms as we decrease
$\theta_\mathrm{max}$ from 90$^\circ$ to 30$^\circ$.   

It is also worthwhile to consider the ``apples-to-oranges''
comparison of the $x_+(x_E)$ curve in \fig{fig:dndxe} and the $x_+$,
$m_g=0$ curve in \fig{fig:dndxmg} (top).  In particular, we observe
that the former has a reduced peak height at $x\sim\mu/E$ but is
larger for higher values of $x$.  This generic behavior is due to a
non-trivial interplay between the Jacobian, 
which always multiplies the $dN_g/dx_+dk_T$ integrand by a number
greater than 1, and the numerically reduced $\kmax=x_EE$ used in \eq{eq:dndxpxe}.  \fig{fig:dndxdktwo} shows that the collinear approximation is better at larger values of $x$ as demonstrated by the small values of $dN_g/dxdk_T$ at $k_T\sim k_\mathrm{max}$ compared to those at lower $k_T$.  The Jacobian, though,
introduces an integrable divergence at $k_T=x_EE$, so the
$x_+(x_E)$ curve is systematically larger than the $x_+$ curve at larger
values of $x$.  Since the total gluon yield, $\langle N_g \rangle$, must
be unmodified by a change in integration variables, the increased
yield at larger $x$ must come at the expense of gluon yield at smaller
values of $x$; hence, the numerical reduction in $k_\mathrm{max}$ wins
in this small $x$ region.  This redistribution of probability means that $\langle x\rangle$ from $x_+(x_E)$ is always greater than $\langle x \rangle$ from $x_+$.  For any given set of medium parameters and path length,
$\langle N_g \rangle$ must be the same for both the $x_+(x_E)$ and $x_+$ implementations; since $\langle x \rangle$ is always larger for the former, using $x_+(x_E)$ will always produce a smaller $R_{AA}$ than $x_+$.  For any given $\theta_\mathrm{max}$, $dN_g/dxdk_T$ with $x=x_+$ is integrated out to a larger $k_\mathrm{max}$ than for $x=x_E$.  Therefore, the $x_+$ interpretation will always yield a smaller $R_{AA}$ than the $x_E$ interpretation.  Using similar reasoning the $x_E$ calculation with $\theta_\mathrm{max}=\pi/2$ will always generate a smaller $R_{AA}$ than the $x_E$ with $\theta_\mathrm{max}=\pi/4$ calculation.  Based on these arguments we expect that the extracted medium parameter $dN_g/dy$ will be ordered from smallest to largest according to $x_+(x_E)$, $x_+$, $x_E$ with $\theta_\mathrm{max}=\pi/2$, $x_E$ with $\theta_\mathrm{max}=\pi/4$. 
The $dN_g/dy$ obtained via the statistical analysis (described below) and listed in \tab{tab:extract} demonstrate exactly this ordering.

\begin{figure}[!htb]
\centering
$\begin{array}{c}
\includegraphics[width=\columnwidth]{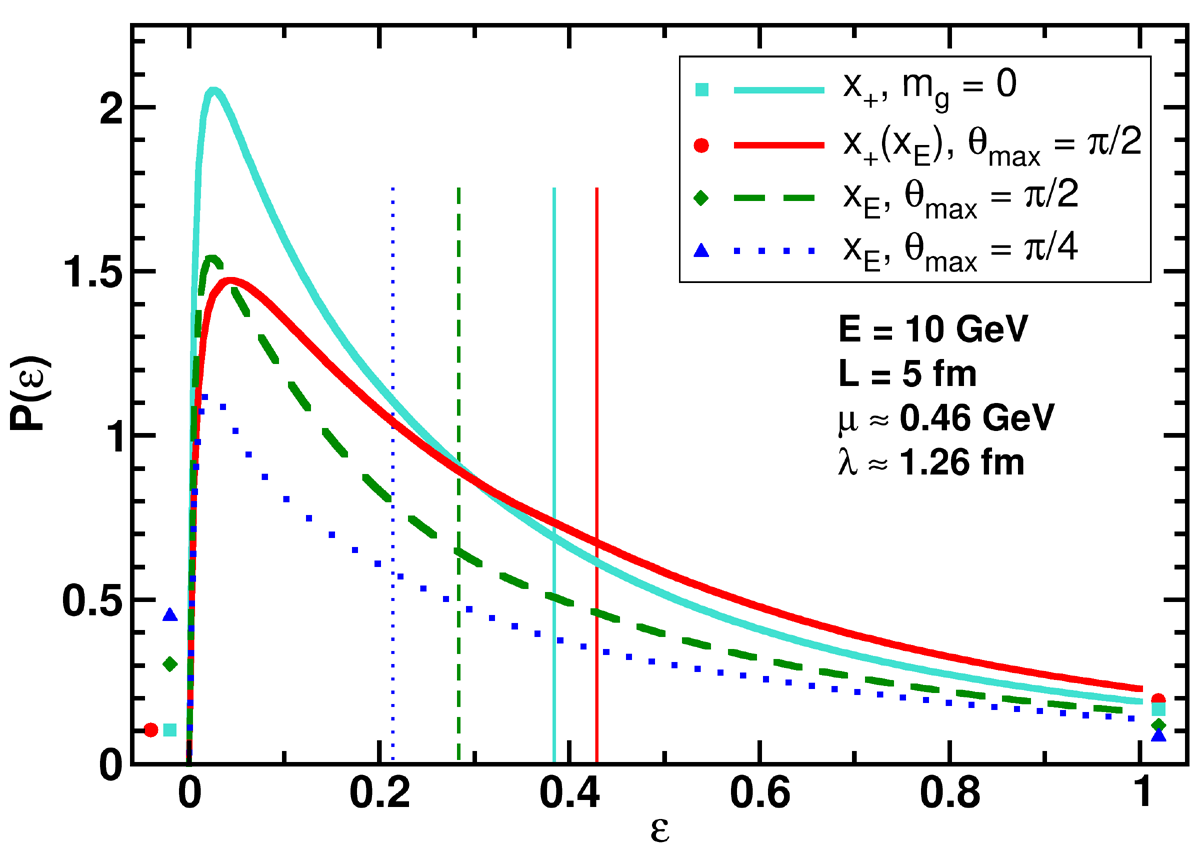} \\
\includegraphics[width=\columnwidth]{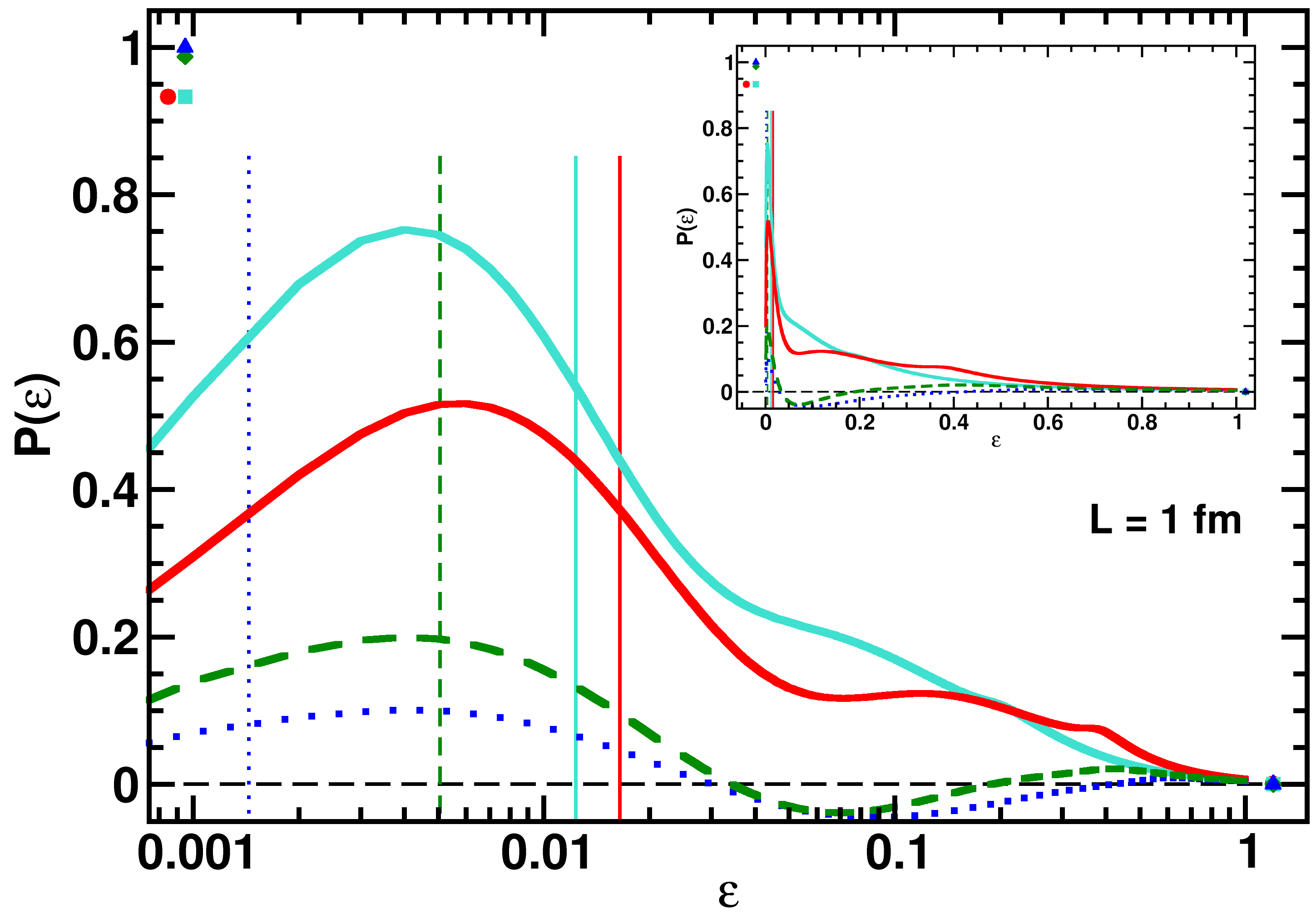}
\end{array}$
\caption{\label{fig:peps}
Plots of $P(\epsilon)$ for a 10~GeV quark in typical RHIC conditions with \mbox{$L=5$~fm} (top) and \mbox{$L=1$~fm} (bottom).  $P(\epsilon)$ obtained by convolving the single inclusive spectra $dN_g/dx$ shown in Figs.~\ref{fig:dndxe} and \ref{fig:dndxmg}.  Vertical lines represent the $\langle\epsilon\rangle$ for each $P(\epsilon)$ distribution.  Symbols represent the weight of the $\delta$ functions at $\epsilon=0$ and $\epsilon=1$.  Inset shows results on a linear scale for clarity.}
\end{figure}

A crucial ingredient for any jet quenching calculation is $P(\epsilon)$, the probability distribution for the fraction of energy, $\epsilon$, radiated by a \highpt parton.  Described above, $P(\epsilon)$ is obtained via a Poisson convolution of the single inclusive radiated gluon spectrum, $dN_g/dx$, assuming the independent emission of the multiple radiated gluons.  $P(\epsilon)$ can be decomposed into discrete and continuous pieces,
\be
P(\epsilon) = P^0\delta(\epsilon)+\tilde{P}(\epsilon)+P^1\delta(1-\epsilon).
\ee
$P^0$ is the probability of radiating no gluons (and hence no energy loss) and is given by
\be
P^0=\exp(-\langle N_g\rangle).
\ee
$P^1$ encodes the probability ``leakage,'' the probability that the radiating parton loses a fraction of energy greater than unity, that results from the assumption of independent emissions used here.  We show in \fig{fig:peps} the $P(\epsilon)$ generated from four of the $dN_g/dx$ curves investigated in this work (themselves displayed in Figs.~\ref{fig:dndxe} and \ref{fig:dndxmg}): $x_+$ with $m_g=0$, $x_+(x_E)$, $x_E$ with $\theta_\mathrm{max}=\pi/2$, and $x_E$ with $\theta_\mathrm{max}=\pi/4$.  The ordering of the average fractional energy loss, $\langle \epsilon \rangle$, (represented by the vertical lines in \fig{fig:peps}) is consistent with the qualitative arguments of the previous paragraph.  
As noted previously $\langle N_g\rangle$ is identical for the $x_+$ and $x_+(x_E)$ models; hence, $P^0$ is also identical for the two.  
Previous evaluations of $P(\epsilon)$ using the $x_+$ interpretation approximated $dN_g/dx_+(x_+)\approx dN_g/dx_E(x_E)$.  We find a systematic, though small, difference between the $\langle \epsilon \rangle$ from this approximation and that resulting from the correct transformation of $dN_g/dx_+$ to $dN_g^J/dx_E$; the modest variation in $P(\epsilon)$ also produces only a $\sim10-20\%$ change in the extracted $dN_g/dy$.  
The continuous parts of $P(\epsilon)$, $\tilde{P}(\epsilon)$, have the expected ordering: from smallest to largest according to $x_E$ with $\theta_\mathrm{max}=\pi/4$, $x_E$ with $\theta_\mathrm{max}=\pi/2$, $x_+$.
This ordering follows simply from the integration of the $dN_g/dxdk_T$ curves out to successively larger $k_\mathrm{max}$.
The non-trivial ordering of $\tilde{P}(\epsilon)$ for the $x_+$ and $x_+(x_E)$ interpretations shows explicitly the redistribution of probability from smaller to larger $\epsilon$ due to the combined effects of the Jacobian, \eq{eq:jacobian}, and the numerically reduced $k_\mathrm{max}$ of \eq{eq:dndxpxe}.  

Ultimately, we wish to quantify the effects on extracted medium
parameters of: (1) the two collinearly equivalent interpretations of
$x$, (2) varying $k_\mathrm{max}$ (via $\theta_\mathrm{max}$), and (3)
the thermal mass of the gluon.  We will
focus here on $\pi^0$ $R_{AA}(\eqnpt)$ measurements obtained from the 5\%
most central Au+Au collisions at $\surd s = 200$ AGeV \cite{Adare:2008cg}.  We follow WHDG \cite{Wicks:2005gt} in the
implementation of energy loss: leading order pQCD-based production
spectra for gluons and light quarks, followed by in-medium energy
loss, and, finally, KKP fragmentation into pions.  Initial state effects
such as Cronin enhancement and shadowing are neglected; as such the
$p_T$ dependence of $R_{AA}$ here is stronger than in works that
better describe the trend of the data \cite{Gyulassy:1999zd}.   

So far in this paper we have only discussed medium-induced gluon
radiation, and for one set of $R_{AA}$ calculations we continue to use
purely radiative energy loss. However, it is known that for pQCD in the
kinematic regimes appropriate to both RHIC and LHC, elastic scattering
leads to a significant contribution to the 
total in-medium energy loss \cite{Mustafa:2004dr,Wicks:2005gt}.  Therefore, we will also calculate $R_{AA}$
using convolved inelastic and collisional loss.  We use the
Braaten-Thoma calculation \cite{Braaten:1991jj,Braaten:1991we} of the
mean elastic energy loss and use the fluctuation-dissipation theorem to
estimate the width of the elastic energy loss distribution \cite{Wicks:2005gt}.  We note
that these assumptions are a poor approximation for RHIC conditions because the number of collisions is typically too small for the central limit theorem to apply.  Hence, Gaussian distributions do not do a good job of representing the actual, highly skewed distributions involved; substantially improved results were derived in
\cite{Wicks:2008}.  However, for the purposes of this paper we are
not concerned with assessing the theoretical uncertainties resulting
from the treatment of the elastic channel; this is a very interesting
question in its own right.   

For the radiative energy loss calculation, we include the full
multi-gluon fluctuations through the Poisson convolution. We also
account for path-length fluctuations due to parent parton production
points and trajectories in the medium for both inelastic and
collisional loss with an approximate implementation of Bjorken expansion \cite{Wicks:2005gt}. A standard Glauber modeling of the medium with a
diffuse Woods-Saxon nuclear density function is used \cite{Wicks:2005gt}. Specifically,
hard production is assumed to scale with binary collisions; the medium
density is assumed to follow participant scaling.  The strong coupling 
constant is held fixed at $\alpha_s=0.3$.  We use thermal field theory
to relate $\mu=gT$ and $\lambda=1/\rho\sigma$, with
$\sigma\propto1/\mu^2$ \cite{Wicks:2005gt}.  The only independent variable left is the
single input parameter $dN_g/dy$, the rapidity density of gluons in
the pure glue QGP assumed here (increasing $dN_g/dy$ increases the
medium density and decreases $R_{AA}$).  The rigorous statistical
analysis of \cite{Adare:2008cg,Adare:2008qa} was then used to
determine the value and 1-$\sigma$ uncertainty of $dN_g/dy$ that yields a theoretical curve that ``best fits'' the data given the experimental statistical and systematic errors.

\begin{figure}[!htb]
\centering
$\begin{array}{l}
\includegraphics[width=\columnwidth]{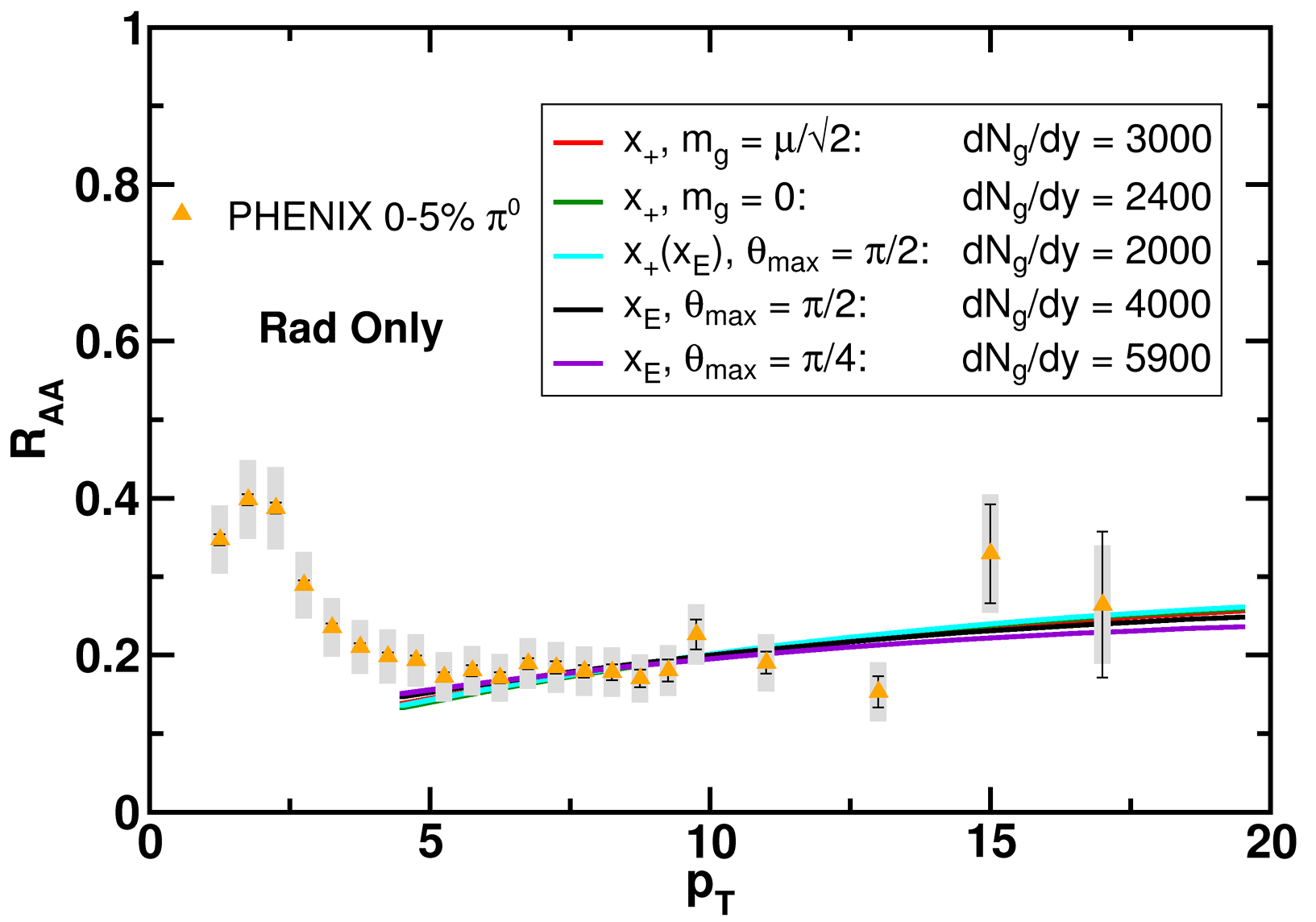} \\
\includegraphics[width=\columnwidth]{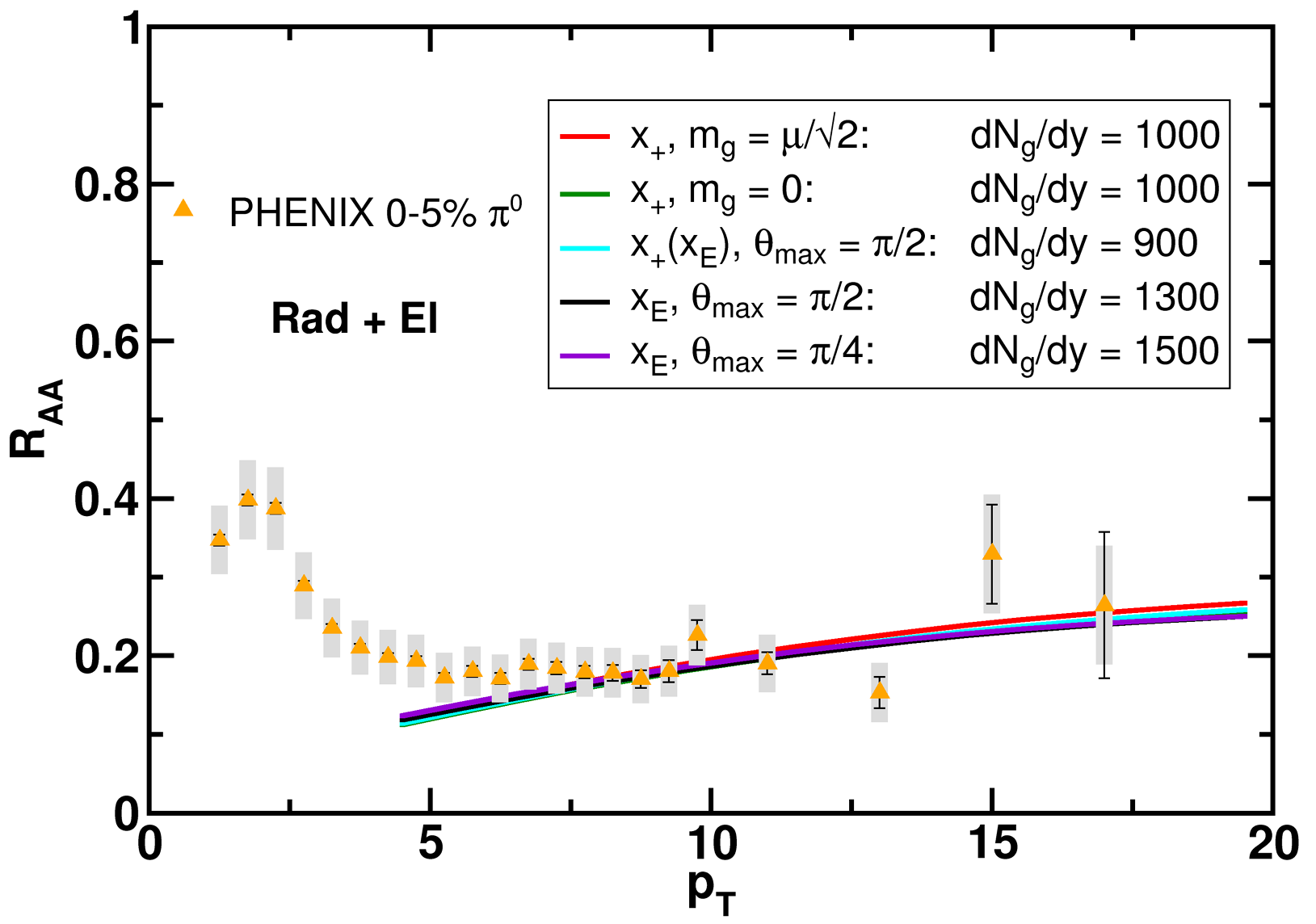}
\end{array}$
\caption{\label{fig:raa}
The ``best fit'' curves of the five models discussed in the text to
PHENIX measurements of the 5\% most central $\pi^0$ $R_{AA}$
\cite{Adare:2008cg}: top--including only radiative energy
loss, bottom-- including both radiative and collisional energy loss. 
Fits performed using the methods of
\cite{Adare:2008cg,Adare:2008qa}. Uncorrelated 
errors are represented by bars whereas correlated errors are shown as
grey boxes, and there is an overall scale uncertainty of $\pm12$\% not
shown. 
}
\end{figure}

We show in \fig{fig:raa} the PHENIX measurement of 
$\pi^0$ \raapt for the 5\% most central collisions
and the best fit curves to the data for the five different
implementations of the radiative energy loss considered in this paper (see
\tab{tab:models}) for both purely radiative energy loss (top)
and convolved radiative and elastic loss (bottom).  
We note that, when compared at similar centrality bins, the STAR measurements of \cite{Abelev:2007ra}
$\pi^++\pi^-$ $R_{AA}$ systematically differ from the shown PHENIX
$\pi^0$ $R_{AA}$ \cite{Adare:2008cg} by as much as
$\sim50\%$.  Such a discrepancy represents a potential source of additional systematic uncertainty in extracted medium properties that we do not attempt to evaluate here.

The results in \fig{fig:raa} indicate that the extracted
$dN_g/dy$ values for the massive ($m_g=\mu/\surd2$) and massless
($m_g=0$) emitted gluon cases are quite similar despite
the large differences in $dN_g/dx$ for large path lengths, \fig{fig:dndxmg} (top).
As we have shown this behavior is due to the interaction of the effects of a thermal gluon mass on the coherence length and the radiation kernel.  
As the path length decreases, the difference between the massive and
massless integrated $dN_g/dx$ decreases until eventually the massive distribution exceeds the massless one.  For central RHIC collisions the results shown in \fig{fig:dndxmg} (bottom) indicate this path length is $\sim1$~fm.  As noted in \cite{Baier:2001yt} physical observables are quite sensitive to the specific numerical choice of the IR cutoff taken when used to approximate the inclusion of a thermal mass for the radiated gluon.  The lack of sensitivity to the particular choice of gluon mass seen in this paper is a pleasant surprise. For the convolved energy loss, the extracted medium
density is consistent for massive and massless radiated gluons.  For identical values of $dN_g/dy$ the massless case yields a smaller $R_{AA}$; that the extracted medium parameter is consistent is likely due to the statistical analysis trading off goodness of fit in the normalization of $R_{AA}$ for slightly different $p_T$ dependencies.

Addressing the primary goal of this paper, we observe from the results
in \fig{fig:raa} that the extracted values of $dN_g/dy$ for the
radiative only energy loss models $x_+(x_E)$; $x_E$,
$\thetamax=\pi/2$; and $x_E$, 
$\thetamax=\pi/4$ vary almost by 200\%.  It cannot be overemphasized
that the first two cases---for which $dN_g/dy$ varies by
$\sim100$\%---are \emph{exactly equivalent} under the collinear
assumption used in the energy loss derivation.  As discussed
previously, for any particular $\theta_\mathrm{max}$, $x_+(x_E)$
produces the smallest \raacomma. Thus, $x_+(x_E)$ with
$\thetamax=\pi/2$ represents a lower bound on \raa for fixed
medium parameters.  Unfortunately,
it is not possible to rigorously define an upper bound to the
theoretical uncertainty on \raa since decreasing
$\theta_\mathrm{max}\rightarrow0$  makes
$R_{AA}\rightarrow1$.  We choose the values of observables calculated from the $x_E$ with $\theta_\mathrm{max}=\pi/4$ implementation
as a working definition of the upper
bound.  
We consider this $\theta_\mathrm{max}$ as representing a reasonable $\mathcal{O}(1)$ variation of
the coefficient of the $k_T$ cutoff that necessarily increases
$R_{AA}$ for a given medium density; the surprise, of course, is just
how sensitive $R_{AA}$ is to such a variation of
the cutoff.  It is worth noting that the larger $dN_g/dy$ values extracted using the radiative energy loss only models are difficult to reconcile with known RHIC $dN_{ch}/dy$ multiplicities \cite{Back:2002uc}.  

\begin{figure}[!htb]
\centering
$\begin{array}{l}
\includegraphics[width=\columnwidth]{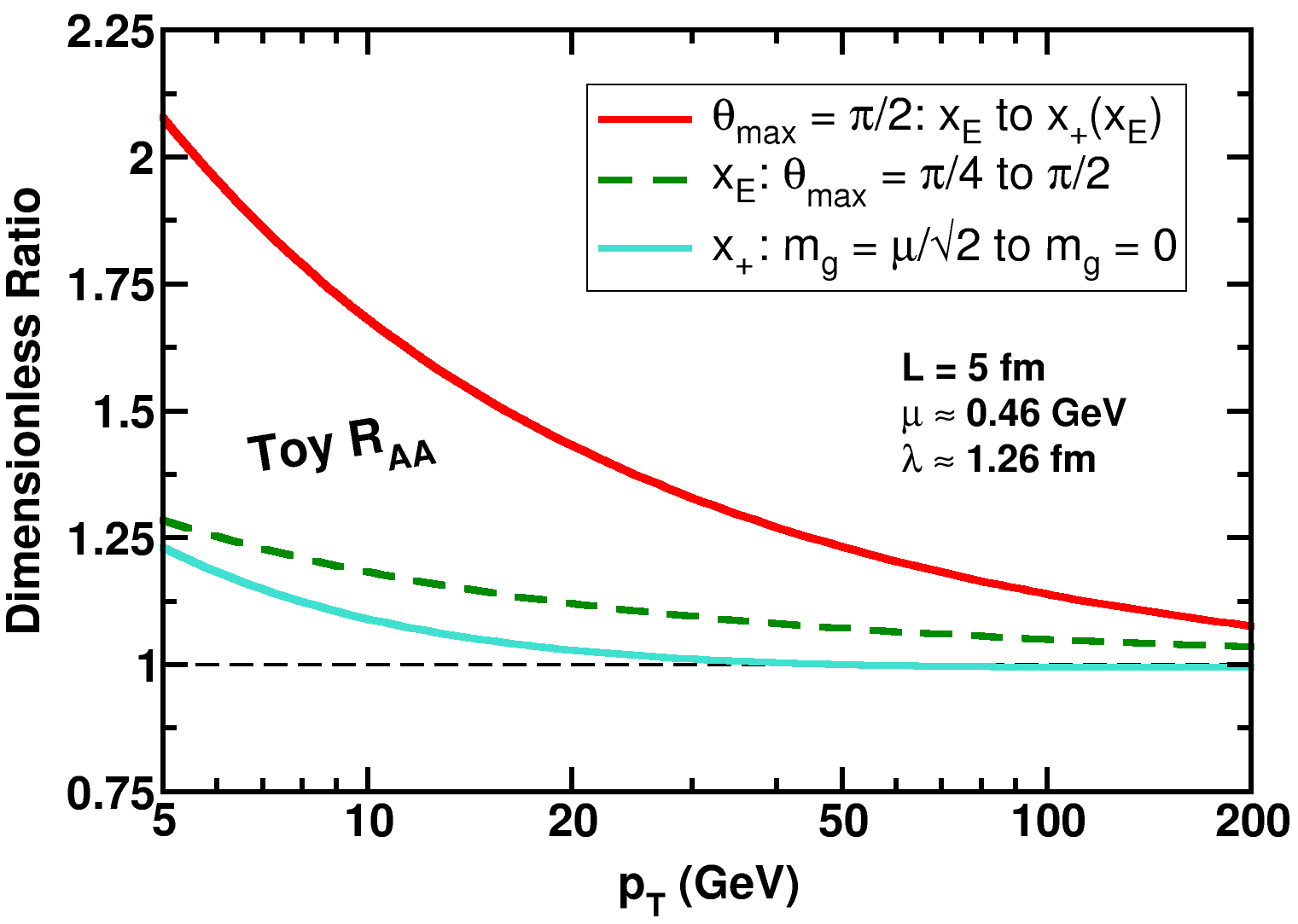} \\
\includegraphics[width=\columnwidth]{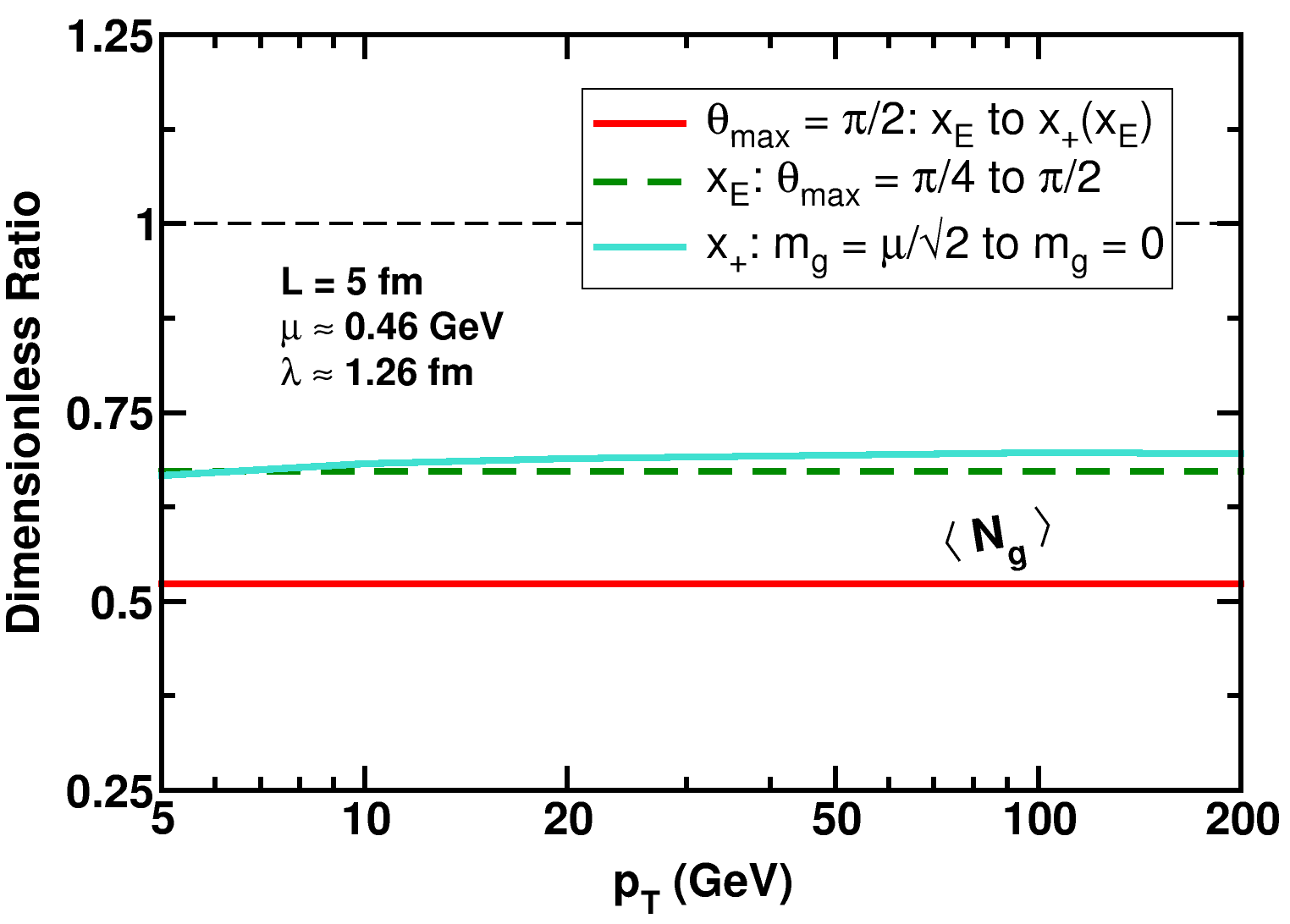}
\end{array}$\caption{\label{fig:endep}
Ratios of a toy $R_{AA}$ model (top) and ratios of the average number of
emitted gluons, $\langle N \rangle$, (bottom) as a function of parent
quark energy for the
five models discussed in the text.  Collinear and gluon mass effects
die out with increasing energy for the toy $R_{AA}$ but do not for
$\langle N \rangle$.  
}
\end{figure}

An important question that must be addressed is whether or not the
sensitivity of extracted medium properties to variations in the
implementation  of the collinear approximation decreases with
increasing radiating parton energy.
To address this question, we show in \fig{fig:endep} a ``toy model''
quark partonic \raapt calculation for $L=5$~fm fixed path length and
typical RHIC medium parameters used above. The sensitivity of the
partonic \raapt to different implementations of the collinear
approximation decreases with increasing parton energy. For example, the
$\simeq100\%$ difference at 5~GeV between the collinearly equivalent
$x_+(x_E)$ and $x_E$ interpretations decreases to $\simeq10\%$ at
200~GeV. But, we observe that the number of emitted gluons, $\langle
N_g \rangle$, remains sensitive to the uncertainties introduced by the
collinear approximation; the $\sim$ factor of 2 difference
in $\langle N_g \rangle$ shown in \fig{fig:endep} is nearly energy
independent.

The convergence of the ratios of \raapt with increasing energy can be
understood at least partly as a consequence of the small $x$ pileup in
$dN_g/dx$.  For an energy loss model with $P(\epsilon)$ and a power
law production spectrum $dN/dp_T\propto p_T^{-(n+1)}$ we have that
$R_{AA}(\eqnpt)\simeq\int(1-\epsilon)^nP(\epsilon;\eqnpt)d\epsilon$.
Then for any two models, say 1 and 2, 
\be
\frac{R_{AA}^1}{R_{AA}^2}(\eqnpt)\simeq 1 - n\Bigl(\bigl\langle\epsilon_1\bigr\rangle(\eqnpt)-\bigl\langle\epsilon_2\bigr\rangle(\eqnpt)\Bigr),
\ee
and the ratio of modification factors automatically approaches 1 as
the mean energy loss decreases with increasing \ptcomma.  On the other
hand, the constancy of $\langle N_g \rangle$ in \pt is due to
the surprising energy independence of the area under the peak in
the $dN_g/dx$ distribution.  As the energy increases, the distribution
becomes more sharply peaked, and its normalization becomes more and
more dominated by this energy-independent area.   

\section{Conclusions}
In our introduction, we laid out the numerous approximations used in
deriving the opacity expansion energy loss formulae.  In one form or
another, these assumptions are at the foundation of the four energy
loss calculations that are frequently compared to RHIC data.  In
particular, these calculations all rely 
on the lowest order term in a collinear expansion, where the
small parameter is $k_T/xE$.  Unfortunately, we have found that the opacity
expansion energy loss kernel $dN_g/dxdk_T$ drastically violates the
collinear approximation for small values of $x$, the region of $x$
most important for computing many observables including the leading
particle suppression and the average number of emitted bremsstrahlung
gluons.  While we did not explicitly check this violation for the
other energy loss formalisms, it is highly likely they also will be similarly
strongly affected by the large angle radiation that, by assumption of
collinearity, is not under theoretical control. 

To leading order in collinearity, the two common definitions of $x$
used in energy loss calculations---$x=x_+$, the fraction of light-cone
plus momentum, and $x=x_E$, the fraction of energy taken away by the
radiated gluon---are equivalent.  We found that for RHIC conditions, the
``best fit'' medium density $dN_g/dy$ extracted using these two
different, but collinearly equivalent, definitions of $x$ varies by a
factor of $\sim2$.   

While one may use the $x_+(x_E)$ interpretation with
$\theta_\mathrm{max}=\pi/2$ as a natural lower bound for the
systematic theoretical uncertainty of $R_{AA}$, the upper bound is
less obvious.  We take the results from the $x_E$ and
$\theta_\mathrm{max}=\pi/4$ as a working definition; in this case the
extracted medium density increases another $\sim50\%$ over the $x_E$
and $\theta_\mathrm{max}=\pi/2$ model.  We note that there is no sense of
a ``central value'' or Gaussian distribution for this uncertainty
band: the rigorous notion of a ``best'' interpretation of $x$ or the
``correct'' UV cutoff for the lowest order collinear results does not
exist prior to a calculation based on a more exact analytic
derivation. 

The uncertainties we quote decrease significantly when elastic energy
loss is included although the effects of uncertainty in the
collisional channel were not considered here.  While the leading
particle $R_{AA}$ appears to suffer less systematic uncertainty at LHC
energies, the uncertainty in other observables, such as the mean
number of emitted gluons, are energy independent.  Finally, the effect
of a thermal gluon mass on the extracted medium density is
surprisingly small due to non-trivial coherence effects.  This last
result implies that the specifics of the short path length energy loss
behavior are very important; future work should, therefore, go beyond
the $L\gg\lambda$ approximation.   

One of the great debates over the past several years has been the
so-called ``discrepancy'' of extracted medium parameters from the four
energy loss models.  An oversimplified description would be that the
density found when comparing AMY, GLV, and HT to data is reasonably
consistent while that found from BDMPS is a factor $\sim2-3$ times
larger \cite{Adare:2008cg,Adare:2008qa,Bass:2008rv}.  It is natural
and right to begin any calculation with 
assumptions about the relevant and irrelevant physics of the problem.
However, that the four models include and exclude vastly different
physics implies to us that any consistency found must be considered
coincidental, even surprising.  We see the path forward not in teasing
out the origins of the extracted difference but in finding observables
that can falsify the basic assumptions about the relevant physics of
the quark gluon plasma.  Is the medium strongly or weakly coupled?
What are its degrees of freedom?  If energy loss is perturbative, are parton interactions with the medium better approximated by many soft scatterings or a few hard ones?  In order to answer these questions 
we must have not only
well-controlled experiments but well-controlled theories.  It is,
thus, imperative that quantitative estimates be made of the systematic
uncertainty introduced into theoretical results from the simplifying 
assumptions---distinct from the physics assumptions---made in the
calculation.  

As discussed previously there is no notion of ``one standard
deviation'' of theoretical uncertainty associated with the collinear
approximation.  Moreover, we have not quantified the consequences of
the collinear approximation for the other three energy loss 
models, although this is both an extremely interesting and clearly
important problem.  Nevertheless, we strongly suspect that should this
uncertainty be quantified for the other models, then the different values of
medium density so far extracted from data would be mutually consistent
within the systematic theoretical uncertainties. 
As we have shown in this work, any hope for a quantitative extraction 
of medium density from \highpt physics at RHIC using the GLV formalism
requires a far more careful treatment of non-collinear radiation; this
is almost certainly true for the AMY, BDMPS-Z-ASW, and HT formalisms,
too. It is interesting that at LHC the leading
particle suppression seems a rather collinearly safe observable
whereas the average number of emitted gluons and the spectrum of soft
gluons is not.  This suggests that---at our current level of
theoretical understanding---leading particle observables and jets
measured using narrower cones may provide more sensitive tests of
jet quenching at the LHC than, e.g., full jet reconstruction with
large cones.
\\
\\
The authors wish to thank Miklos Gyulassy, Yuri Kovchegov, and the members of the \textsc{techqm} Collaboration for valuable discussions; in particular we thank Ulrich Heinz and Urs Wiedemann for their help in elucidating the differences between the $x_+$ and $x_E$ definitions.  Additionally the authors thank Jamie Nagle for quantitatively extracting $dN_g/dy$ from PHENIX data and the calculations presented in this work.  We also thank Ulrich Heinz for reading and commenting on the manuscript.  WAH thanks the Aspen Center for Physics for support during his stay. This work was supported by the Office of Nuclear Physics in the Office of Science of the U.S.\ Department of Energy under Grant Nos.\ DE-FG02-05ER41377 and DE-FG02-86ER40281.

\def\eprinttmppp@#1arXiv:@{#1}
\providecommand{\arxivlink[1]}{\href{http://arxiv.org/abs/#1}{arXiv:#1}}
\def\eprinttmp@#1arXiv:#2 [#3]#4@{\ifthenelse{\equal{#3}{x}}{\ifthenelse{
\equal{#1}{}}{\arxivlink{\eprinttmppp@#2@}}{\arxivlink{#1}}}{\arxivlink{#2}
  [#3]}}
\providecommand{\eprintlink}[1]{\eprinttmp@#1arXiv: [x]@}
\renewcommand{\eprint}[1]{\eprintlink{#1}}
\providecommand{\adsurl}[1]{\href{#1}{ADS}}
\renewcommand{\bibinfo}[2]{\ifthenelse{\equal{#1}{isbn}}{\href{http://cosmolog%
ist.info/ISBN/#2}{#2}}{#2}}

\end{document}